\documentclass[english,eqsecnum,prd,aps,nofootinbib,superscriptaddress,longbibliography,tightenlines,11pt]{revtex4-2}

\usepackage{makecell}

\usepackage{array,multirow,graphicx}
\usepackage{enumerate}
\usepackage{comment}
\usepackage{graphicx}
\usepackage[table]{xcolor}
\usepackage{euscript,amssymb}
\usepackage{mathrsfs}
\usepackage{setspace}
\usepackage{makecell}
\usepackage{amsfonts}
\usepackage{enumitem}
\usepackage{amsmath}
\usepackage{amsbsy}
\usepackage{mathrsfs}
\usepackage{fancyhdr}
\usepackage{esint}
\usepackage{amsthm}
\usepackage{empheq}
\usepackage{mathtools}
\usepackage{amsbsy}
\usepackage{epsfig}
\usepackage{color}

\usepackage{booktabs}
\usepackage{colordvi}

\DeclareFontFamily{OT1}{pzc}{}
\DeclareFontShape{OT1}{pzc}{m}{it}{<-> s * [1.2] pzcmi7t}{}
\DeclareMathAlphabet{\mathpzc}{OT1}{pzc}{m}{it}

\usepackage[unicode=true, pdfusetitle,
bookmarks=true,bookmarksnumbered=false,bookmarksopen=false,
breaklinks=false,pdfborder={0 0 1},backref=false,colorlinks=false]
{hyperref}
\usepackage{graphicx}

\usepackage{xcolor,fontawesome}
\usepackage{amsfonts}
\usepackage{enumitem}
\usepackage{amsmath}
\usepackage{amssymb}
\usepackage{colordvi}
\usepackage{fancyhdr}
\usepackage{esint}
\usepackage{colordvi}

\begin{document}

\title{Role of Gauss-Bonnet corrections in a DGP brane gravitational collapse}

\author{Yaser Tavakoli}
\email{yaser.tavakoli@guilan.ac.ir} 
\affiliation{Department of Physics,
University of Guilan, Namjoo Blv.,
41335-1914 Rasht, Iran}
\affiliation{School of Astronomy, Institute for Research in Fundamental Sciences (IPM), P.O. Box 19395-5531 Tehran, Iran}

\author{Ahad K. Ardabili}
\email{ahad.ardabili@altinbas.edu.tr} 
\affiliation{Department of Basic Sciences, Alt{\i}nba\c{s} University, 34217 Istanbul, Turkey} 

\author{Mariam Bouhmadi-L\'{o}pez}
\email{mariam.bouhmadi@ehu.eus}
\affiliation{Ikerbasque, Basque Foundation for Science, 48011 Bilbao, Spain}
\affiliation{Department of Physics, University of the Basque Country UPV/EHU, Bilbao 48080, Spain}
\affiliation{EHU Quantum Center, University of the Basque Country UPV/EHU}

\author{Paulo Vargas Moniz}
\email{pmoniz@ubi.pt} 
\affiliation{CMA-UBI and Departamento de F\'{\i}sica, Universidade da Beira Interior, 6200 Covilh\~{a}, Portugal}

\date{\today}


\begin{abstract}
 An Oppenheimer-Snyder (OS)-type collapse is considered for a Dvali-Gabadadze-Porrati (DGP) brane, whereas a Gauss-Bonnet (GB) term is provided for the  bulk. 
We  study the combined  effect of the DGP induced gravity plus the GB curvature, regarding any modification of the general relativistic  OS  dynamics. Our  paper has a twofold objective. On the one hand,  we  investigate the nature of singularities that may arise at the collapse end state. It is shown that all dynamical scenarios for the  contracting brane would end 
in  one of the following cases, depending on conditions  imposed:  either a central shell-focusing singularity or what we designate as a ``sudden collapse singularity.'' On the other hand, we also study  the deviations of the exterior spacetime from the  standard Schwarzschild  geometry, which emerges in our modified OS scenario. 
Our purpose is to investigate whether a black hole always 
forms regarding this brane world model. We find situations where a naked singularity emerges instead. 
 
\end{abstract}

\date{\today}


\maketitle

\section{Introduction}

One of the most impressive features in relativistic astrophysics is the occurrence of continual gravitational collapse of supermassive stars due to their own gravitational fields.
This provides a  framework toward understanding the basic mechanism for the formation of  white dwarfs, neutron stars, then possibly  supermassive  black holes \cite{Joshi:1987wg,Joshi:2008zz}.
In particular, one of the main consequences of  general relativistic gravitational collapse is the formation of central singularities at the collapsing final state, where the gravitational field  and, consequently, the spacetime scalar curvature are predicted to diverge. 

An important inquiry that  may arise in this context is whether there exist circumstances that avoid the formation of the black hole event horizon prior to the occurrence of the central singularity, allowing it to be visible to the external world. 
It turns out that, despite the theoretical predictions for the existence of naked singularity solutions in general relativity (GR), Penrose argued, through his cosmic censorship conjecture    \cite{Penrose:1969pc}, that such situation would not occur: central  singularities should be hidden within event horizons and therefore cannot be observed from the rest of the spacetime. 
The previous appraisal notwithstanding, at a more fundamental level it is
believed that a singularity is indeed not a location where quantities really become infinite, but instead where GR merely breaks down. Thereby, GR should be superseded by a robust singularity-free gravitational theory (as yet unknown, quantum theory of gravity). In the past decades,  there has been enormous efforts in order to put forward a more complete theory for gravitation, which would be free of such obstacles.

Higher-dimensional brane world models \cite{Deffayet:2000uy,Dvali:2000hr,Cho:2001nf,Randall:1999ee,Randall:1999vf}, due to the presence of both bulk and brane
curvature terms in their actions, have a different gravitational behavior from their general relativistic counterpart \cite{Kofinas:2003rz,Deffayet:2001pu}. They provide interesting extensions of our parameter space for gravitational theories. In particular, within  cosmological and astrophysical contexts, brane world models admit spacetime singularities of rather unusual form and nature \cite{Brown:2005ug,BouhmadiLopez:2008nf,BouhmadiLopez:2011xi,BouhmadiLopez:2009jk}. Among these models, there are scenarios of specific interest. On the one hand, we have  brane gravity proposed by  Dvali {\em et al.} (known as the DGP induced-gravity model), in which the brane action constitutes  an alteration (with respect to the Einstein-Hilbert term for GR),  so that it provides a 
low energy (or infrared) modification through an induced-gravity term \cite{Dvali:2000hr}.
On the other hand, we can also consider  Gauss-Bonnet (GB) (higher-order curvature) corrections, which are introduced 
for the  bulk  \cite{Charmousis:2002rc,Kim:2000pz,Nojiri:2002hz,Davis:2002gn,Gravanis:2002wy}. So, whereas DGP features play at  infrared range, the GB  elements change gravity at high energy (or ultraviolet). These two additions allow us to broaden the scope of research with respect to gravitational collapse, beyond and encompassing GR as a suitable limit, possibly 
suggesting a peculiar phenomenology from the quantum gravity sector. 

The previous paragraphs set the context of our herewith 
reported research, with the one just above closely assisting in defining our purposes. Let us be more precise. 
The gravitational collapse of a stellar object  was first studied by Oppenheimer and Snyder \cite{Oppenheimer:1939ue} in the context of GR.
By considering a homogeneous spherically symmetric matter cloud with vanishing internal pressures and zero rotation, they showed  that the cloud collapses simultaneously to a (central) spacetime singularity, covered within a black hole event horizon.
This mechanism has been extended later to more general models of gravitational collapse in  GR   (see \cite{Joshi:2008zz,Tavakoli:2014dfa} and references therein) and alternative  theories of gravity \cite{Maeda:2006pm,Jhingan:2010zz,Tavakoli:2015llr,Ziaie:2019klz}, namely, brane induced gravity \cite{Bruni:2001fd,Gannouji:2012wk} including GB corrections \cite{Tavakoli:2020xgc},   as well as in quantum gravity approaches \cite{Bojowald:2005qw,Bojowald:2008ja,Tippett:2011hz,Bambi:2013caa,Barcelo:2007yk,Hajicek:1992mx,Marto:2013soa,Tavakoli:2013rna,Tavakoli:2013tpa,Vaz:2001mb, Parvizi:2021ekr}. 

Further regarding  our setup, let us mention that a supermassive star can both be  large  and have huge energy density. That motivates and suggests that  it should be of specific interest to investigate, for instance,  how the combination of DGP brane gravity with GB terms can alter the final state, 
possibly with  asymptotic states rather different from those within a  GR framework. Therefore, questions such as how the influence of higher-dimensional  effects would modify the evolution of event horizons and formation of central  singularity are sufficient to enthuse us to investigate  gravitational collapse as described above. 
In this sense, it has been shown that a collapse  in the DGP-GB system can avoid the formation of a black hole. Concretely, if one considers a large scale system (the initial matter configuration assumed to be distributed within a large radius) that corresponds to a marginally bound [$k = 0$, here $k$ is a constant in Friedmann-Lema\^itre-Robertson-Walker (FLRW) metric which represents the curvature of the space] case, then the collapse outcome depends on the initial mass of the system, and a ``naked sudden singularity" can emerge before the collapse process reaches the vanishing physical radius $R=0$ point; see Ref. \cite{Tavakoli:2020xgc} for details. In our present paper, we develop instead from a  generalized initial configuration which has been assumed to include the bound collapse ($k=1$), and, as it will be discussed in the rest of the paper, new plausible scenarios have been obtained.

Thus, our current paper has a twofold objective. On the one hand, we  study the  effect of the DGP brane world model, containing a GB curvature in the bulk, regarding any modification of the general relativistic  Oppenheimer-Snyder (OS) \cite{Oppenheimer:1939ue} collapse. Concretely, we  investigate the nature of singularities that may arise at the collapse end state. We find that, depending on conditions that we will  make explicit, two type of singularities can happen: a central shell-focusing singularity or  what we designate as a  sudden collapse singularity (SCS). In addition, our other purpose is 
to identify any deviation of the exterior spacetime from the standard Schwarzschild black hole, which emerges in the relativistic OS model. 

Motivated as we described in the previous paragraphs, we thus organized our paper as follows. In Sec.~\ref{model}, we will present our model of gravitational collapse, by considering a closed FLRW for the contracting brane, containing a dust fluid as matter content, with a GB term in the bulk. We will then present the evolution equation of the collapse on the brane and provide different classes of solutions. 
In Sec.~\ref{singularity}, we discuss the physically relevant solutions for the evolution of the contracting brane and present the emergence of a specific type of singularity at which, for a nonzero physical radius,  the energy density of the brane remains finite, while the time derivative of the Hubble rate diverges.\footnote{In a cosmological setting, such event is called a ``sudden singularity'' \cite{Barrow:2004xh}  and it  can occur in a late time universe, filled with or without phantom matter, within a finite cosmic time.}
In Sec.~\ref{exterior}, the matching condition at the boundary of the collapsing object will be investigated and  possible solutions to the exterior geometry at the final state of the collapse will be determined. Finally, in Sec.~\ref{conclusion}, we present our conclusion and a discussion of our work.

\section{Gravitational collapse in the DGP-GB model}
\label{model}

We consider the DGP brane world model in which
a GB curvature term is present in the five-dimensional Minkowski bulk containing a FLRW brane with an induced-gravity term.
Moreover, all the energy-momentum is localized on the four-dimensional brane. Therefore, the total gravitational action can be written as 
\begin{align}
S &= \frac{1}{2} \int dx^5 \sqrt{- g^{(5)}} \, M^3_{(5)} \left[{\cal R}_{(5)}  +\alpha\mathcal{L}_{\rm GB}\right] + \frac{1}{2} \int dx^4 \sqrt{- g} \, M^2_{(4)} \left({\cal R}_{(4)} + {\cal L}_{\rm matt} \right),
\label{action}
\end{align}
where $\mathcal{L}_{\rm GB}$ is the Lagrangian density of the GB sector
\begin{align}
\mathcal{L}_{\rm GB} =  {\cal R}_{(5)}^2  - 4{\cal R}_{(5)ab}{\cal R}_{(5)}^{ab}    +  {\cal R}_{(5)abcd}{\cal R}_{(5)}^{abcd}.
\end{align}
Here, $g^{(5)}$ and $g$ denote, respectively, {\em determinants} of the metric in the bulk and  the induced metric on the brane.
The couplings $M_{(5)}^{-3}=\kappa^2_{(5)}$ and $M_{(4)}^{-2}=\kappa^2_{(4)}$ are respectively, the four- and five-dimensional gravitational constants. Moreover, $\alpha>0$ is the GB coupling constant. The standard DGP model is the special case with $\alpha=0$ and possesses a crossover scale, defined by $r_c=\kappa^2_{(5)}/2\kappa^2_{(4)}$, which marks the transition from the four-dimensional brane to the five-dimensional bulk.\footnote{Note that, the cross over scale, $r_c$, considered here has an extra factor of $1/2$ comparing to   that defined in \cite{Brown:2005ug}, but the physics does not change, i.e. when substituting the definition of  $r_c$ given in \cite{Brown:2005ug} and that of our herein definition in terms of $\kappa_5^2$ and $\kappa_4^2$, the Friedmann equation (\ref{modifiedfriedmann}) will be the same.}

We consider an OS model for  gravitational collapse on the brane. It consists of a spacetime manifold $\mathbb{R}\times\Sigma$ on the brane, where $\mathbb{R}$ represents the time direction and $\Sigma$ is a homogeneous and isotropic three-manifold bearing a dust cloud. The homogeneity and isotropy  imply that $\Sigma$ must be a maximally symmetric space, so the space is spherically symmetric. 
Then, the spacetime geometry within this spherically symmetric  cloud is described by the {\em comoving} coordinates $(t, r, \theta, \varphi)$ \cite{Oppenheimer:1939ue}
\begin{eqnarray}
ds^2=-dt^2 + \frac{a^2(t)}{\left(1+kr^2/4\right)^2}\Big[dr^2+r^2d\Omega^2\Big],
\end{eqnarray}
where $t$ is the proper time, with $a(t)$ being the scale factor of the collapse, and $d\Omega^2$ is the line element of the standard unit two-sphere. Moreover, $R$ is denoted to 
be the physical radius, given by
\begin{eqnarray}
R(t, r) = \frac{ra(t)}{1+kr^2/4}\, =:\, \mathbf{r} a(t)\, .
\label{Radius}
\end{eqnarray}
In GR, the case $k=0$ describes a marginally bound collapse, in which shells at infinity have zero initial velocity [i.e., $\dot{R}(t_0, r)=0$]. The case in which $k=-1$ states an  
unbound collapse whose shells at infinity have positive initial velocity [$\dot{R}(t_0, r)>0$]. Finally, the case $k=+1$ represents a bound collapse for which shells at infinity have negative initial velocity [$\dot{R}(t_0, r)<0$].

The generalized Friedmann equation of a spatially closed (i.e., with $k=+1$) DGP brane with a GB term in a Minkowski bulk [defined in Eq.~(\ref{action})]
reads \cite{Kofinas:2003rz}
\begin{equation}
h^2=\frac{\kappa_{(5)}^2}{6r_c}\rho+
\frac{\epsilon}{r_c}\left(1+\frac83\alpha h^2\right)h,
\label{modifiedfriedmann}
\end{equation}
where, $h$ is defined in terms of the %Hubble 
parameter $H$ of the brane as 
\begin{eqnarray}
h^2 \ := \ H^2+\frac{k}{a^2}\, .
\label{Hubble-origin}
\end{eqnarray} 
For the case $\alpha = 0$ and $k=0$, Eq.~(\ref{modifiedfriedmann}) reduces to the Friedmann equation in the standard DGP model \cite{Deffayet:2000uy}. The parameter $\epsilon$ can take the values $\epsilon=\pm 1$;
the $(-)$ sign for $\epsilon$ can be visualized as the interior of the space bounded by a hyperboloid (the brane), whereas the $(+)$ sign corresponds to the exterior.\footnote{In the DGP brane cosmological model, the $(+)$ sign for $\epsilon$ corresponds to the {\em self-accelerating}  branch  and the $(-)$ sign corresponds to the {\em normal} branch.}
Moreover,  $\rho$ stands for the total energy density of the
brane. For an evolving brane, containing a perfect fluid with an equation of state parameter $w=\gamma-1$, which represents  a collapsing star, 
the energy density is well described by
\begin{equation}
\rho=\rho_{0}\left(\frac{a_{0}}{a}\right)^{3\gamma} \, ,
\label{mattercontent}
\end{equation}
where $\gamma,\rho_{0}$ and $a_{0}$ are constants.\footnote{Quantities with the	subscript $0$ denote their values at the initial configuration of the collapsing star.} 
To compare our model to an OS-type collapse, we are interested in a (pressureless) dust fluid, so henceforth we will set $\gamma=1$.

It is convenient to  introduce the following
dimensionless variables in order to analyze  the evolution of the Hubble rate as a function of the  energy density of the brane \cite{BouhmadiLopez:2008nf}:
\begin{eqnarray}
\bar h  := \frac83 \frac{\alpha}{r_c}h , \quad  \bar\rho  := \frac{32}{27}
\frac{\kappa_{(5)}^2\alpha^2}{r_c^3} \rho , \quad  b := \frac83\frac{\alpha}{r_c^2} \, . \quad  \label{eq8a}
\end{eqnarray}
Using  these new  variables, the evolution equation (\ref{modifiedfriedmann}) for the $(-)$ and $(+)$ branches become
\begin{subequations}
	\label{Friedmann-tot}
	\begin{eqnarray}
	&& {\bar h}^3+{\bar h}^2+b\bar h-\bar\rho=0 \quad  {\rm for~ the~ (-)~ branch}, \label{Friedmannnb}  \\
	\nonumber\\
	&& {\bar h}^3-{\bar h}^2+b\bar h+\bar\rho=0 \quad   {\rm for~the~ (+)~ branch}. \quad 
	\label{Friedmannnb2}
	\end{eqnarray}
\end{subequations}
By solving these equations for $\bar{h}$, we can obtain the solutions to the Hubble rate as
\begin{eqnarray}
\bar{H} \, = \, -\sqrt{\bar{h}^2- \frac{\bar{k}^2}{a^2}} \, ,
\label{Hubble0}
\end{eqnarray} 
where we have defined
\begin{equation}
\bar{H}\, :=\, \frac83 \frac{\alpha}{r_c}H = br_cH\, , \quad {\rm and} \quad \bar{k}^2\, :=\, kb^2r_c^2\, .
\end{equation}
We note that  $\bar{H}$ above is always real and negative when the argument under the square root is positive. 
The two cubic equations (\ref{Friedmannnb}) and (\ref{Friedmannnb2}) can be transformed to one another by changing the parameter therein as  $\bar{h}\rightarrow -\bar{h}$. Thus, positive solutions of the first equation correspond to the negative solutions of the second one, and vice versa. Interestingly, the solutions $\bar{h}$ in the expression  
(\ref{Hubble0}) under the square root is of quadratic form, which indicates that  considering either of the branches leads to  identical solutions for the Hubble rate \cite{BouhmadiLopez:2009jk}. Therefore,  we will henceforth consider only the first equation [the one corresponding to the $(-)$ branch] and will present the solutions of the second equation only briefly in  Table \ref{plus-Branch}. Nevertheless, we should notice that, as we will explain carefully in what follows, having the same solution for $\bar{h}$ does not necessarily lead to the same end state for the collapse. 
In fact, other elements will also play an important role as we  will explain.

Let assume that $\bar{k}>0$. Then, at the initial configuration, positiveness of the argument under the square root implies that  $|\bar{h}(t_0)|$ should be slightly larger than $\bar{k}/a_0$, where the collapsing cloud has a maximum size at $a(t_0)=a_0$.  Since the scale factor decreases as the collapse evolves, then in order to keep the argument under the square root positive,  the parameter $|\bar{h}|$ should increase toward $a=0$ (and identically zero physical radius, $R=0$). This means that $|\bar{h}|$  has a minimum value at the initial  hypersurface, i.e., $|\bar{h}_0|=|\bar{h}|_{\rm min}$. This minimum should satisfy the initial condition
\begin{eqnarray}
\bar{h}_{0} \gtrsim \frac{\bar{k}}{a_0}\, \quad \quad {\rm or} \quad \quad \bar{h}_{0} \lesssim -\frac{\bar{k}}{a_0}\, .
\label{initial-condition}
\end{eqnarray}
Then, for all $t>t_0$, the physically reasonable solutions for $\bar{h}(t)$ are those that  hold the condition  
\begin{equation}
|\bar{h}(t)|\geq \frac{\bar{k}}{a(t)} \, .
\label{condition-h}
\end{equation}

In order to determine the number of real roots of the cubic equations (\ref{Friedmann-tot}) it is helpful to define\footnote{The evolution equation (\ref{Friedmann-tot}) will be solved analytically following the method introduced in Ref.~\cite{BouhmadiLopez:2008nf}.}  the discriminant function $N$ 
as \cite{Abramowitz:1965,BouhmadiLopez:2009jk}
\begin{equation}
N=Q^3+S^2,
\label{eq10}
\end{equation}
where $Q$ and $S$ are
\begin{equation}
Q:=\frac{1}{3}\left(b-\frac{1}{3}\right),
\quad \quad  S:=\frac{1}{6}b+\frac{1}{2}\bar{\rho}-\frac{1}{27} \, .
\label{eq11}
\end{equation}
For  later convenience, in  analyzing  the number of physical solutions of the modified Friedmann equation, we rewrite $N$ as \cite{BouhmadiLopez:2009jk}
\begin{equation}
N=\frac{1}{4}(\bar{\rho}-\bar{\rho}_{A})(\bar{\rho}-\bar{\rho}_{B}),
\label{eq12}
\end{equation}
where
\begin{subequations}
	\begin{eqnarray}
	\bar{\rho}_{A}\, &:=&\, \frac{2}{27}\left[1+\sqrt{(1-3b)^3}\right]-\frac{b}{3}\, ,
	\label{rhoone}
	\\
	\bar{\rho}_{B}\, &:=&\, \frac{2}{27}\left[1-\sqrt{(1-3b)^3}\right]-\frac{b}{3}\, .
	\label{rhotwo}
	\end{eqnarray}
\end{subequations}
The number of roots of Eqs.~(\ref{Friedmann-tot}) is determined by the sign of $N$:  If $N$ is positive, there exists a unique real root. If
$N$ is negative, there are three real solutions, and for vanishing $N$, all roots are real and at least two are equal.

From Eq.~(\ref{eq12}) it is clear that the sign of $N$ depends on the value of the energy density of the brane $\bar\rho$ with respect to the parameters $\bar{\rho}_{A}$ and $\bar{\rho}_{B}$. Moreover, $\bar{\rho}_{A}$ and $\bar{\rho}_{B}$ are defined in terms of the parameter $b$. Then, in order to determine the sign of $N$ and, consequently, the number of solutions of the modified Friedmann equations for the $(-)$  and $(+)$  branches, it is convenient to distinguish four ranges for the values of $b$, as (A)   the range $0<b<\frac{1}{4}$, (B) the range $\frac{1}{4}\leq b<\frac{1}{3}$, (C) the value  $b=\frac{1}{3}$, and, finally, (D) the range $b>\frac{1}{3}$. 

In the following subsections, we will study the possible solutions of the Friedmann equations (\ref{Friedmann-tot}) for different ranges of $b$, namely A--D.

%%%%%%%%%%%%%%%%%%%%%%%%%%%%%%%%%%%%%%%%%%%%%%%
\subsection{The case $0<b<\frac{1}{4}$}
\label{A}
%%%%%%%%%%%%%%%%%%%%%%%%%%%%%%%%%%%%%%%%%%%%%%%

In this case, the values of $\bar{\rho}_{A}$ and $\bar{\rho}_{B}$  are real and,
furthermore, $\bar{\rho}_{A}>0$ and $\bar{\rho}_{B}<0$.
Then, the number of real roots of the cubic equation (\ref{Friedmannnb}) depend on the minimum energy density $\rho_0$ (of the dust fluid) on the brane. Let us define the dimensionless  energy density $\bar\rho_0$ of the collapsing fluid  as
\begin{eqnarray}
\bar\rho_0 \,  :=\,  \frac{32}{27}
\frac{\kappa_{(5)}^2\alpha^2}{r_c^3} \rho_0\, =\,  \frac{\kappa_{(4)}^2}{3} b^2r_c^2
\rho_0 .
\label{energy-trans}
\end{eqnarray}
Then,  different situations for the roots can be distinguished within  two cases: $\bar\rho_0>\bar{\rho}_A$ and $\bar\rho_0\leq \bar{\rho}_A$. In what follows in this subsection we will analyze the possible real solutions of (\ref{Friedmannnb}) for these two initial conditions.
 
%\vspace{2mm}

%%%%%%%%%%%%%%%%%%%%%%%%%%%%%%%%%%%%%%%%%%%%%%%
\subsubsection{The  condition $\bar\rho_0>\bar{\rho}_A$}
%%%%%%%%%%%%%%%%%%%%%%%%%%%%%%%%%%%%%%%%%%%%%%%

For this choice of initial condition at any $t>t_0$ in the future, the condition $\bar{\rho}>\bar{\rho}_A$ will always hold because, in the collapse process, the energy density of the dust is incremental as it moves from the initial configuration toward $R=0$.
Thus, the function $N$ will  always remain positive, thereby, there exists a  unique real solution for (\ref{Friedmannnb}),
\begin{equation}
\bar{h}_1(\eta) = \frac{1}{3}\left[2\sqrt{1-3b}\cosh\left(\frac{\eta}{3}\right)-1\right],
\label{EXP1}
\end{equation}
where the parameter $\eta$  is defined as
\begin{equation}
\quad  \cosh(\eta) \coloneqq \frac{S}{\sqrt{-Q^3}}~, \quad \quad \sinh(\eta) \coloneqq
\sqrt{-\frac{N}{Q^3}} \, .
\label{eq16}
\end{equation}
Using the first relation in  Eq.~(\ref{eq16}), we
can find an expression for the energy density of the brane in terms of $\eta$ as
\begin{eqnarray}
\bar{\rho}_1(\eta)\, =\, \frac{2}{27}\left[1+\sqrt{(1-3b)^3}\cosh(\eta)\right]-\frac{b}{3}\, . \quad \quad 
\label{density1}
\end{eqnarray}
From $\bar{\rho}_1(\eta)\geq\bar{\rho}_0>\bar{\rho}_A$, it follows that the parameter $\eta$ is constrained  by  $\cosh(\eta)\geq\cosh(\eta_0)>\cosh(\eta_A)$. This yields 
a lower bound on  $\eta$ as $\eta\geq\eta_0>\eta_A$, where  $\eta_A$ and $\eta_0$ are given by
\begin{subequations}
	\begin{eqnarray}
	\cosh(\eta_A)\, &=&\, \frac{9b+27\bar{\rho}_A-2}{2\sqrt{(1-3b)^3}}  \, , \label{eta-A}\\
	\cosh(\eta_0)\, &=&\, \frac{9b+27\bar{\rho}_0-2}{2\sqrt{(1-3b)^3}}  \,
	. \label{eta-0}
	\end{eqnarray}
\end{subequations}
%%%
Now, Eq.~(\ref{eta-A}) yields $\eta_A=0$, thereby, the solution $\bar{h}_1(\eta_A)\equiv \bar{h}_{A}$ is determined as
\begin{equation}
\bar{h}_{A} \coloneqq \frac{1}{3}\left[2\sqrt{1-3b}-1\right].
\label{EXP1-A}
\end{equation}
Likewise, by using the energy density (\ref{density1}) into the Eq.~(\ref{mattercontent})  the scale factor is obtained  in terms of   $\eta$ as
\begin{eqnarray}
a_1(\eta) &=& a_{0}\left(\frac{\bar{\rho}_0}{\bar{\rho}_1(\eta)}\right)^\frac{1}{3}. \quad \qquad 
\label{scale-1}
\end{eqnarray}
As $\eta$ increases,   $\bar{\rho}_1(\eta)$ increases as well so the scale factor decreases toward zero, where $\eta$ diverges.

\begin{figure}
	\begin{center}
		\includegraphics[scale=0.55]{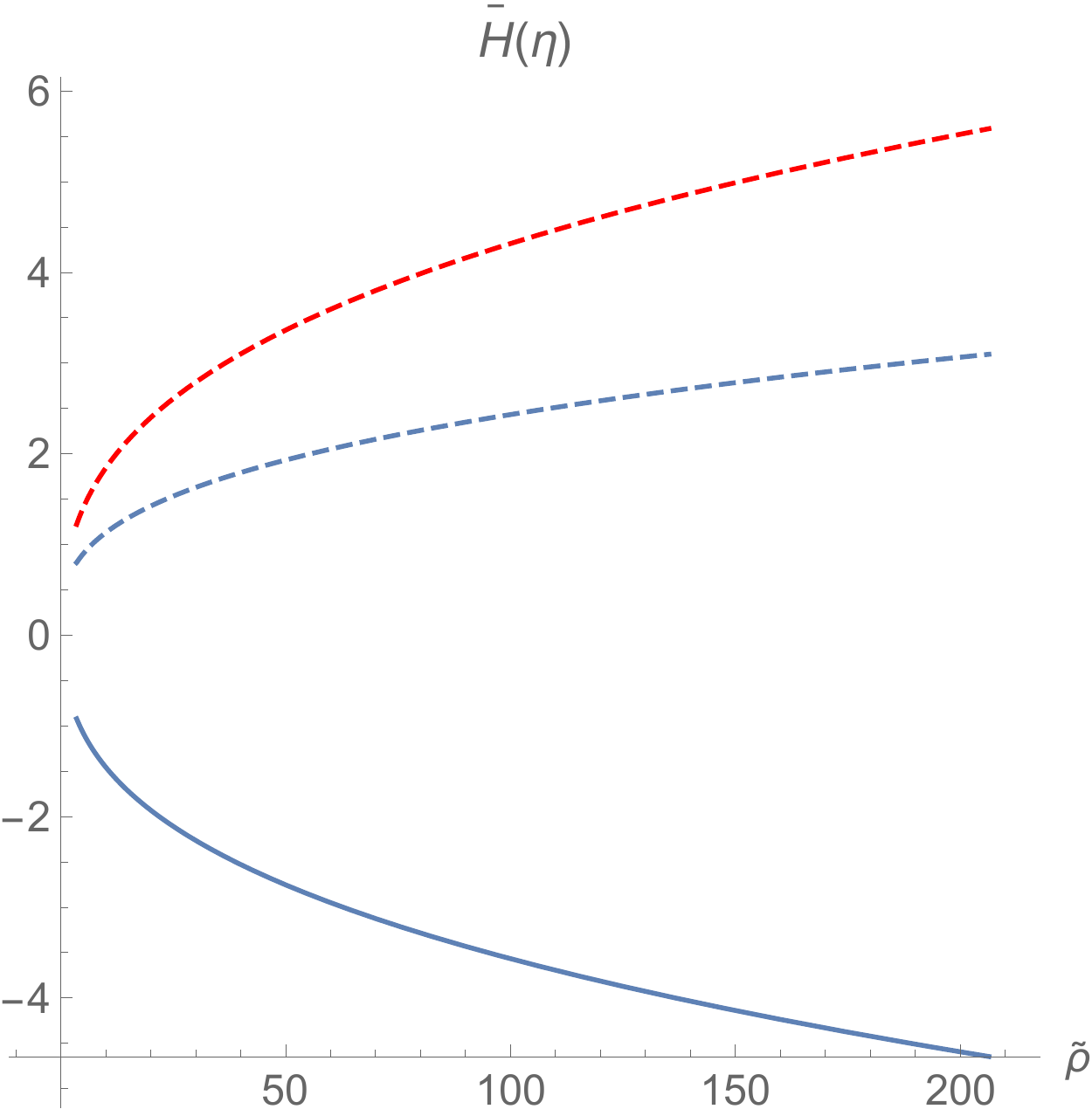} 
		\caption{The evolution of the (dimensionless)  $\bar{H}_1(\eta)$ in terms of $\bar\rho$, in the range  of parameters $0<b<1/4$ and for the initial condition $\bar{\rho}_0>\bar{\rho}_A$: it is shown that for all $\eta>\eta_{\rm min}$,  the first term, $\bar{h}_1(\eta)$, in  Eq.~(\ref{h-bar-1}) (dashed red curve)  increases  faster than the second term, $\bar{k}/a_1(\eta)$ (dashed blue curve), so  $\bar{H}_1(\eta)$ (solid blue curve) blows up (with a minus sign) when $\bar\rho_1(\eta)$ tends to infinity.}\label{Case0}
	\end{center}
\end{figure}

The condition (\ref{condition-h}) implies that, since the scale factor $a_1(t)$ decreases by time as collapse proceeds, the solution $\bar{h}_1(\eta)$ should increase much faster than $\bar{k}/a_1(\eta)$.
It turns out that, since $\bar{h}_1(\eta)$ was initially positive valued, it should remain positive for all times in the future.
By imposing the initial condition (\ref{initial-condition}) into the solution (\ref{EXP1}), this produces a more restricted range for the parameter $\eta$; it is
 $\eta_0>\eta_{\rm min}>0$, where  $\eta_{\rm min}$ is given  by
\begin{align}
\cosh\left(\frac{\eta_{\rm min}}{3}\right) \coloneqq \frac{3(\bar{k}/a_0) +1}{2\sqrt{1-3b}}\, .
\end{align}
On the other hand,  $\bar{h}_1(\eta)$ should belong to the range
\begin{equation}
\bar{h}_1(\eta_0) > \bar{h}_1(\eta) > \bar{h}_{\rm 1 min}\, ,
\label{EXP1-b}
\end{equation}
where, $\bar{h}_{\rm 1min}\equiv \bar{k}/a_0$.
Thus, $\bar{H}_1(\eta)$ is determined by setting Eqs.~(\ref{EXP1})  and (\ref{scale-1}) into Eq.~(\ref{Hubble0}):
\begin{eqnarray}
\bar{H}_1(\eta) \, = \, -\sqrt{\bar{h}_1^2(\eta)- \frac{\bar{k}^2}{a_1^2(\eta)}} \, .
\label{h-bar-1}
\end{eqnarray} 
Because both terms under the square root of 
Eq.~(\ref{h-bar-1}) increase from the initial state, we  expect two possible scenarios for the evolution of the collapse. The first is a case in which the term $\bar{h}_1(\eta)$  increases faster than the second term $\bar{k}/a_1(\eta)$, so that, as the collapse progresses toward a ``centering'' point,  $\bar{H}_1$ blows up. The other scenario is that the second term in Eq.~(\ref{h-bar-1}) increases faster than the first term so that at some moment in the future, the  two terms cancel each other and $\bar{H}_1$ vanishes.
Nevertheless, the numerical analysis  shown in Fig.~\ref{Case0}  indicates that initially,  when the  collapse starts [i.e., $\bar{H}_1(t_0)\lesssim0$] and satisfies the condition (\ref{EXP1-b}),  this condition will continue to hold through the whole collapsing process. Therefore, as the collapse proceeds, the first term in Eq.~(\ref{h-bar-1}) increases faster than the second term so that, $\bar{H}_1(\eta)$ will blow up (with a minus sign) at the vicinity of $a=0$. 

\vspace{2mm}

%%%%%%%%%%%%%%%%%%%%%%%%%%%
\subsubsection{The  condition $\bar\rho_0\leq \bar{\rho}_A$}

For this choice of initial condition, the number of solutions of Eq.~(\ref{Friedmannnb}) will depend on the values of the energy density $\bar\rho$ with respect to $\bar{\rho}_{A}$. 
As the  energy density   blueshifts in time (i.e., $\rho$ increases as the collapse evolves), we can distinguish three regimes: (i) a {\em low energy} regime, where $\bar{\rho}<\bar{\rho}_{A}$, (ii) a {\em limiting} regime with
$\bar{\rho}=\bar{\rho}_{A}$, and 
(iii) a {\em high energy} regime for which $\bar{\rho}>\bar{\rho}_{A}$. In what follows, we will  summarize the solutions of Eq.~(\ref{Friedmannnb})  and their properties through these three regimes:
%%%%%%%%%%%%%%%%%%%%%%%%%%%%%%%%%%%%%%%%%%%%%%%%%%%%%
\begin{enumerate}[label=(\roman*)]

\item The low energy regime is given by the range of the fluid energy density  $\bar{\rho}<\bar{\rho}_{A}$, which corresponds to the range of physical radius $R(t)>R_A$, where
\begin{eqnarray}
R_A \coloneqq R_0 \left(\frac{\bar{\rho}_0}{\bar{\rho}_A}\right)^{\frac{1}{3}}
\label{sudden-radius-1}
\end{eqnarray}
is the physical radius, $R_A=\mathbf{r}a_A$ of the collapse in the limit $\bar\rho=\bar{\rho}_A$. Likewise, $R_0=\mathbf{r}a_0$ is the initial value for the physical radius. 
Hereon, the function $N$ is negative, and there are three {\em real}  solutions for $\bar{h}$. Two of these solutions are negative given by
\begin{equation}
\quad \quad  \bar{h}_{2}(\theta)=-\frac{1}{3}\left[2\sqrt{1-3b}\cos\left(\frac{\pi +\theta}{3}\right)+1\right],\label{lowenergy1}
\end{equation}
and
\begin{equation}
\quad \quad  \bar{h}_{3}(\theta)=-\frac{1}{3}\left[2\sqrt{1-3b}\cos\left(\frac{\pi -\theta}{3}\right)+1\right],\label{lowenergy2}
\end{equation}
in which, $0<\theta \leq\theta_0$ defined by
\begin{equation}
\quad \quad  \cos(\theta) \coloneqq \frac{S}{\sqrt{-Q^3}}\, , \quad \quad \sin(\theta) \coloneqq \sqrt{\frac{N}{Q^3}}\, .\label{theta}
\end{equation}
Here, again $\theta_{0}$ is  a parameter at  the initial configuration for which $\bar{\rho}(\theta_0)\equiv\bar{\rho}_0$,
\begin{eqnarray}
\cos(\theta_0) &=& \frac{9b+27\bar{\rho}_0-2}{2\sqrt{(1-3b)^3}}  \, .
\label{energy0}
\end{eqnarray}
From the first relation in Eq.~(\ref{theta}) we obtain the energy density of the dust as
\begin{eqnarray}
\bar{\rho}_2(\theta)\, =\, \frac{2}{27}\left[1+\sqrt{(1-3b)^3}\cos(\theta)\right]-\frac{b}{3}\, .  \ 
\label{density2}
\end{eqnarray}
It is clear that $\bar{\rho}_2(\theta)$ is bounded because of the bound on the {\em cosine} term. Moreover,   since $\bar{\rho}_2(\theta)$ increases as the collapse evolves, so $\bar{\rho}_0\leq\bar{\rho}_2(\theta)<\bar{\rho}_A$. Thus, the upper bound on the energy density is the density $\bar{\rho}_A$, in the limiting regime, which corresponds to the parameter $\theta_A=0$.
This yields a limit on $\theta$ as $0<\theta \leq \theta_0$ or, accordingly, $\cos(\theta_0)\leq \cos(\theta)<1$, as mentioned above.

In the limit  $\theta\rightarrow 0$  where $\bar{\rho}\rightarrow\bar{\rho}_{A}$,  both solutions (\ref{lowenergy1}) and (\ref{lowenergy2})  coincide and  approach a  limiting solution  [cf. next item, Eq.~(\ref{limitingone})],
\begin{eqnarray}
\bar{h}_{2},\ \bar{h}_{3}\ \longrightarrow\ \bar{h}_{A}^\prime \coloneqq -\frac{1}{3}\left[\sqrt{1-3b}+1\right].
\end{eqnarray}
This illustrates a collapse scenario  on the brane that starts to contract from  an initial configuration with $\theta_0$ and $\bar{\rho}_0$, and then by increasing $\bar{\rho}_2(\theta)$, it proceeds toward the limiting regime as $\theta\rightarrow0$. 
In the meantime, for the collapse to  start  in this regime, the initial  $\bar{H}(t_0)$  should be slightly negative at the initial state. The condition (\ref{initial-condition}) in this case reads $\theta_0<\theta_{\rm m}$, where $\theta_{\rm m}$ is defined in
\begin{equation}
\quad \cos\left(\frac{\pi +\theta_{\rm m}}{3}\right)\ \coloneqq\ \frac{(3\bar{k}/a_0)-1}{2\sqrt{1-3b}}\, .
\end{equation}
To have a physical reasonable situation for the collapse, $|\bar{h}(\theta)|$ should increase from the initial configuration as $\bar\rho$ increases.
A numerical analysis of the solutions (\ref{lowenergy1}) and (\ref{lowenergy2}) in  Fig.~\ref{h-123} (cf. blue curves) indicates that the only possible solution describing this situation  is $\bar{h}_2(\theta)$ given by  Eq.~(\ref{lowenergy1}). Indeed, the solution  (\ref{lowenergy2})  cannot be a  physically relevant possibility because it represents an evolution in which $|\bar{h}_3(\theta)|$  decreases as $\bar\rho(\theta)$ grows.
\begin{figure*}
	\begin{center}
		\includegraphics[scale=0.22]{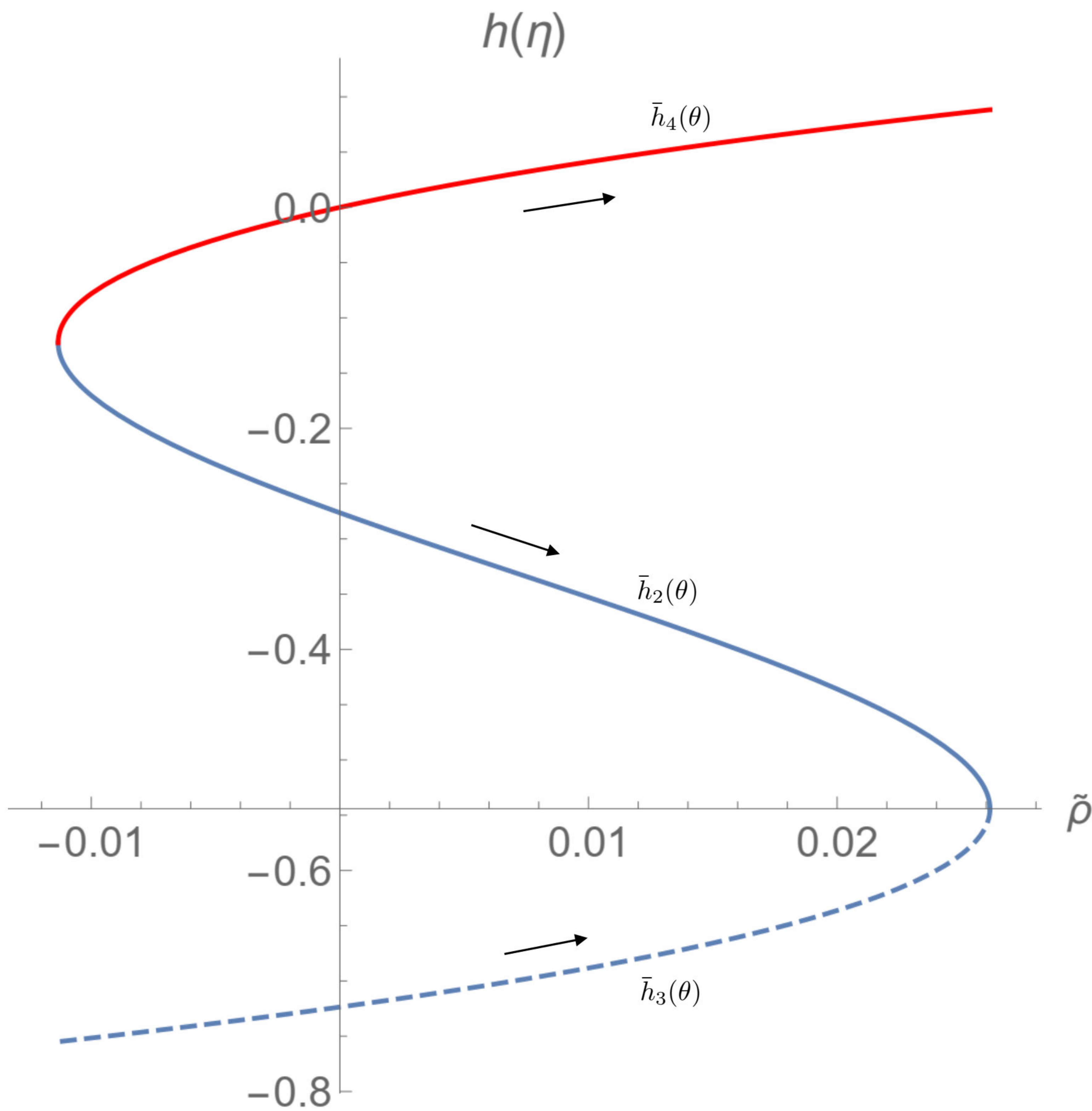}
		\caption{The evolution of  $\bar{h}_2(\theta)$ (blue solid curve), $\bar{h}_3(\theta)$ (blue dashed curve),  and $\bar{h}_4(\theta)$ (red solid curve) given, respectively, by Eqs.~(\ref{lowenergy1}), (\ref{lowenergy2}), and (\ref{lowenergy3}) with the  assumed ranges  of parameters $0<b<1/4$ and $0<\theta<\theta_0$. 
		This was plotted by setting $b=1/8$ and $\theta_0=0.9\pi$ within the allowed regions of parameters.}\label{h-123}
	\end{center}
\end{figure*}

The third  solution in this regime is positive, given by
\begin{equation}
\quad\quad  \bar{h}_{4}(\theta)=\frac{1}{3}\left[2\sqrt{1-3b}\cos\left(\frac{\theta}{3}\right)-1\right],
\label{lowenergy3}
\end{equation}
where $0<\theta\leq\theta_{0}$, 
during which the  quantity $\bar{h}_4(\theta)$  increases from the initial state at $\theta_0$ toward the limiting state at $\theta=0$ (cf. the red solid curve in Fig.~\ref{h-123}). Note that the initial condition for  $\bar{H}_4(\theta)$, namely,  the relation (\ref{condition-h}), implies that $\theta_0<\tilde{\theta}_{\rm m}$, where $\tilde{\theta}_{\rm m}$ is defined by  
\begin{equation}
\cos\left(\frac{\tilde{\theta}_{\rm m}}{3}\right)\, \coloneqq\, \frac{(3\bar{k}/a_0)+1}{2\sqrt{1-3b}}\, .
\label{theta-max-2}
\end{equation} 
When $\theta$ tends to zero [limiting regime; cf item \ref{limiting-item} below], then  $\bar{h}_4(\theta)\rightarrow \bar{h}_{A}$, where $\bar{h}_{A}$ is given by Eq.~(\ref{EXP1-A}).

%%%%%%%%%%%%%%%%%%%%%%%%%%%%%%%%%%%%%%%%%%%%%%%%%%%%%%%%%%%%
\item \label{limiting-item}  In the limiting regime for which $\bar{\rho}=\bar{\rho}_{A}$, the
function $N$ vanishes, so there are two real solutions,
\begin{equation}
\bar{h}_{A}=\frac{1}{3}\left(2\sqrt{1-3b}-1\right),
\label{limitingone}
\end{equation}
and
\begin{equation}
\bar{h}_{A}^\prime\ =\ -\frac{1}{3}\left(\sqrt{1-3b}+1\right).
\label{limitingtwo}
\end{equation} 
The above solutions are the upper bound for the solutions we have found earlier in the low energy regime, when $\theta\rightarrow0$. On the contrary,  the solution (\ref{EXP1}), in the case $\bar\rho_0>\bar\rho_A$,  has its lower bound at $\bar{h}_{A}$; that is,  $\bar{h}_1(\eta)>\bar{h}_{A}$, where $\eta>\eta_A$. 
Indeed, the solution $\bar{h}_1(\eta)$ can be classified in the high energy regime,  presented in the following item.

\item During the high energy regime, the energy density of the brane has a lower bound at $\bar{\rho}_{A}$. 
Since $N>0$ in this case, 
Eq.~(\ref{Friedmannnb}) has a unique solution that  is described by the solution (\ref{EXP1}).
The contracting brane in this regime displays a  (negatively)  decreasing   $\bar{H}_{1}(\eta)$, given by Eq.~(\ref{h-bar-1}),  for the contracting brane that  is proceeding toward the center, $R=0$. 
Therefore, if the collapse enters this regime, it can end up with a  {\em shell-focusing} singularity.
\end{enumerate}
 
To summarize this subsection, let us point out the aspects that are assisting our investigation and that will be 
employed further. In particular, the   reasonable physical solutions  within this range of the $b$ parameter  is the brane contracting solutions described by 
Eqs.~(\ref{EXP1}), (\ref{lowenergy1})  and (\ref{lowenergy3}). In  Secs. \ref{B}--\ref{D}, we will work out solutions for the remaining range of the  $b$ parameter. In the subsequent sections, we will study the type of singularities  (namely,  shell-focusing and  sudden collapse singularities) that  result from the solutions found so far in this section, and those presented  for the remaining range of the $b$ parameter in the rest of this section.

\subsection{The case $\frac{1}{4}\leq b<\frac{1}{3}$}
\label{B}

In this case, $\bar{\rho}_A\leq 0$ and $\bar{\rho}_B< 0$ for which  $N>0$. Since $\bar\rho_0>0$,  there exists a unique real solution for $\bar{h}$ which is described by   Eq.~(\ref{EXP1}) and leads to the 
rate (\ref{h-bar-1}) concerning the   evolution of the brane. 
This solution corresponds to a contracting  brane with a permanent 
increase in $\bar{H}_1(\eta)$,  so that it can blow up at  $R=0$.

\subsection{The case $b=\frac{1}{3}$}
\label{C}

This case corresponds to a unique real solution because $\bar{\rho}_A=\bar{\rho}_B=-1/27<0$ so that $N>0$. The dimensionless quantity $\bar{h}$ in this case reads
\begin{equation}
\bar{h}_5(t)\ =\ \frac{1}{3}\left[(1+27\bar{\rho})^{1/3}-1\right].
\label{b=1/3case}
\end{equation}
This solution is positive and should satisfy the initial condition (\ref{condition-h}) to present a physically relevant contracting  brane. This implies a constraint on the initial energy density of the brane as $\bar{\rho}_0>\bar{\rho}_{\rm min}$, where
\begin{equation}
\bar{\rho}_{\rm min} = \frac{1}{27}\left(\frac{3\bar{k}}{a_0}+1\right)^3 -\frac{1}{27}\, .
\end{equation}
The quantity $\bar{h}_5$ is an increasing function of $\bar{\rho}$. 
Then, $\bar{H}_5$ provided by this solution is a decreasing (with negative sign) function,  so that it blows up as the center is approached.

\subsection{The case $b>\frac{1}{3}$}
\label{D}

In this case, two parameters $\bar{\rho}_A$ and $\bar{\rho}_B$ are complex conjugates. Therefore, $N$ is always positive and Eq.~(\ref{Friedmannnb}) has a unique real solution for $\bar{h}$. This solution reads 
\begin{equation}
\bar{h}_6(\vartheta) =\frac{1}{3}\left[2\sqrt{3b-1}\sinh\left(\frac{\vartheta}{3}\right)-1\right],
\label{Hubble-extra1}
\end{equation}
where, $\vartheta$  is defined by
\begin{equation}
\quad  \sinh(\vartheta):= \frac{S}{\sqrt{Q^3}}~, \quad \quad \cosh(\vartheta):=
\sqrt{\frac{N}{Q^3}} \, .
\label{Hubble-extra2}
\end{equation}
The first relation in Eq.~(\ref{Hubble-extra2}) yields the energy density of the dust as
\begin{eqnarray}
\bar{\rho}_6(\vartheta)\, =\, \frac{2}{27}\left[1+\sqrt{(3b-1)^3}\sinh(\vartheta)\right]-\frac{b}{3}\,,  \quad 
\label{density3}
\end{eqnarray}
which is an increasing function of $\vartheta$ from a convenient initial parameter $\vartheta_0$.

The solution (\ref{Hubble-extra1}) describes a collapsing scenario that begins with an initial  energy density  $\bar\rho_0=\bar\rho_6(\vartheta_0)$ toward  $R=0$.  By setting this solution in Eq.~(\ref{Hubble0}), we obtain the collapse rate $\bar H_6(\vartheta)$, which is a negative decreasing function  within the range of parameter $\vartheta\geq\vartheta_0>\vartheta_{\rm m}$, where $\vartheta_{\rm m}$ is defined by
\begin{equation}
\sinh\left(\frac{\vartheta_{\rm m}}{3}\right) \coloneqq \frac{(3\bar{k}/a_0)+1}{2\sqrt{3b-1}}\, .
\end{equation}
Therefore, as the collapse approaches the center $R=0$, the energy density and $\bar{h}_{6}$  diverge.

\vspace{1cm}

So far in Secs. \ref{A}--\ref{D}, we have analyzed the solution $\bar{h}$ of the cubic equation (\ref{Friedmannnb}) for the $(-)$ branch. So, let us recall that the solutions of the cubic equation (\ref{Friedmannnb2}) can be expressed as the negative  of the solutions of (\ref{Friedmannnb}). In other words, having determined the solutions $\bar{h}$ of the $(-)$ branch we can find the solutions of the $(+)$ branch, simply, by setting $\bar{h} \rightarrow -\bar{h}$.
We have summarized this in Table \ref{plus-Branch}, 
namely, the solutions for the gravitational collapse on the $(+)$ branch of the DGP-GB model. 
The $\bar H$ is equivalent for both $(+)$ and $(-)$ branches, since the solutions $\bar h$ used in the relation  (\ref{Hubble0}) are  quadratic, which means that regardless of  the sign of $\bar h$, the value of $\bar H$ is also determined.  In the next section, we will examine these solutions to find out what type of singularity [shell-focusing singularity (SFS) or SCS] is the result. Furthermore, we will check their compatibility with the energy conditions. To see if the singularity is hidden behind an event horizon, we will study the exterior geometry of the collapsing system.

%%%%%%%%%%%%%%%%%%%%%%%%%%%%%%%%%%%%%%%%%%%%%%%%%%
\begin{center}
	\begin{table*}[ht!]
		\begin{tabular}{  ccccc }
			\Xhline{2\arrayrulewidth}
			\hline\hline \vspace{-2mm}\\
			$b$  & \multicolumn{1}{c}{\begin{tabular}[c]{@{}c@{}} { Initial} \\   { conditions} \end{tabular} }    &  $\bar \rho$   &~~ \qquad   {Solutions}~ $\bar h$ \qquad ~~ & \multicolumn{1}{c}{\begin{tabular}[c]{@{}c@{}}  { Ranges of  } \\   {\bf $\eta, \theta, \vartheta$ } \end{tabular}  }    \vspace{1mm} \\ 
			\hline\\
			%%-------------------------------------------------------------------------------
			$0<b<\frac{1}{4}$ & $\bar\rho_0>\bar\rho_{A}$ & $\bar\rho>\bar\rho_A $ &  ~~~ $\bar{\mathpzc{h}}_1(\eta)=-\frac{1}{3}\left[2\sqrt{1-3b}\cosh\left(\frac{\eta}{3}\right)-1\right]
			$   &   $\eta>0$     \vspace{5mm}\\
			%%-------------------------------------------------------------------------------
			& $\bar\rho_0\leq \bar\rho_{A}$ &   $\bar\rho<\bar\rho_A$ &  ~~~$\bar{\mathpzc{h}}_{2}(\theta)=\frac{1}{3}\left[2\sqrt{1-3b}\cos\left(\frac{\pi +\theta}{3}\right)+1\right]$   &   $0<\theta<\theta_0$    \vspace{2mm}\\
			%%-------------------------------------------------------------------------------
			&  &     & ~~~$\bar{\mathpzc{h}}_{3}(\theta)=\frac{1}{3}\left[2\sqrt{1-3b}\cos\left(\frac{\pi -\theta}{3}\right)+1\right]$   &   $0<\theta<\theta_0$      \vspace{2mm}\\
			%%-------------------------------------------------------------------------------
			&  &     & ~~~$\bar{\mathpzc{h}}_{4}(\theta)=-\frac{1}{3}\left[2\sqrt{1-3b}\cos\left(\frac{\theta}{3}\right)-1\right]$   &    $0<\theta<\theta_0$   \vspace{3mm}\\
			%%=====================================================================
			& $\bar\rho_0\leq \bar\rho_{A}$  &   $\bar\rho=\bar\rho_A$ &    ~~~$\bar{\mathpzc{h}}_{A}=-\bar{h}_{A}=\frac{1}{3}\left[\sqrt{1-3b}+1\right]$   &         \vspace{2mm}\\
			%%-------------------------------------------------------------------------------
			&  &     & ~~~$\bar{\mathpzc{h}}_{A}^\prime=-\bar{h}_{A}^\prime=-\frac{1}{3}\left[2\sqrt{1-3b}-1\right]$   &          \vspace{3mm}\\
			%%================================================================
			$\frac{1}{4}\leq b<\frac{1}{3}$ &  $\bar\rho_0>0$ & $\bar\rho_A,\bar\rho_{B}<0$ &   ~~~$\bar{\mathpzc{h}}_1(\eta)=-\frac{1}{3}\left[2\sqrt{1-3b}\cosh\left(\frac{\eta}{3}\right)-1\right]$   &   $\eta>0$     \vspace{3mm}\\
			%%===============================================================
			$ b=\frac{1}{3}$ &   $\bar\rho_0>0$ & $\bar\rho_A,\bar\rho_{B}<0$ &   ~~~$\bar{\mathpzc{h}}_5 = -\frac{1}{3}\left[(1+27\bar{\rho})^{1/3}-1\right]$   &          \vspace{3mm}\\
			%%-------------------------------------------------------------------------------
			$b>\frac{1}{3}$ &    $\bar\rho_0>0$   & \multicolumn{1}{c}{~\begin{tabular}[c]{@{}c@{}} $\bar\rho_A, \bar\rho_{B}$ {\em complex} \\   {\em conjugates} \end{tabular} } &  ~~~$\bar{\mathpzc{h}}_6(\vartheta) =-\frac{1}{3}\left[2\sqrt{3b-1}\sinh\left(\frac{\vartheta}{3}\right)-1\right]$   &  $\vartheta\geq\vartheta_0$      \vspace{2mm}\\
			%%-------------------------------------------------------------------------
			\hline\hline\Xhline{2\arrayrulewidth}
		\end{tabular}
		\caption{Solutions for $\bar h$ in  gravitational collapse on the  $(+)$ branch of DGP-GB model. Note that all the parameters $\bar{h}, b, N, \bar{\rho}, \eta, \theta, \vartheta$ are the same as defined in Sec.~\ref{model}.}
		\label{plus-Branch}
	\end{table*}
\end{center}

%%%%%%%%%%%%%%%%%%%%%%%%%%%%%%%%%%%%%%%%%%%%%%%%
\section{Nature of the  singularity}
\label{singularity}

In this section, we  study the fate of the  collapse for the various solutions we have found in the previous section  for appropriate  ranges of the parameter $b$. In particular, we will find which of the solutions obtained in the previous section can lead to a SCS by checking the time derivative of $\bar{H}$. Furthermore, we will examine their compatibility with the energy conditions.

\subsection{Collapse: Possible outcomes}

We consider  the  brane filled with a dust fluid collapsing toward the center, $R=0$. The conservation equation for the energy density $\rho$ of the brane reads
\begin{equation}
\dot{\rho}+3H\rho=0.
\label{eq23}
\end{equation}
Since  $\dot\rho>0$, the energy density of the brane
increases from  $\rho_0$ as the collapse proceeds toward $R=0$, i.e., the point where the  relativistic singularity is located.
Consequently, $H$  increases (with negative sign) through the contracting solutions we obtained for different ranges of  $b$ (cf. the previous section). If the brane matter content  blueshifts during its contraction, then  $H$, its time derivative $\dot{H}$, and the energy density $\rho$ of the brane keep increasing until they diverge at the center. In this case, the collapse would end up with a central, shell-focusing singularity.
However,   through a careful analysis of the evolution of $\dot{H}$,  it is demonstrated that  a rather different outcome for the collapse end state is inevitable.
As we will show in the current section, the GB term modifies the evolution of $H$ and its time derivative $\dot{H}$ in such a way that a different outcome, in contrast to  the conventional collapse scenarios, will be plausible.

Let us therefore proceed. By differentiating  Eq.~(\ref{Hubble-origin}) with respect to the comoving time, we get  the time derivative $\dot H$ as
\begin{eqnarray}
\dot{H} = \sqrt{1 + \frac{k}{a^2H^2}} \ \dot{h} + \frac{k}{a^2}\, .
\end{eqnarray}
In above equation, $\dot{h}$ can be obtained by taking the (comoving) time derivative of $h(t)$ in the modified Friedmann equation (\ref{modifiedfriedmann}),
\begin{equation}
\dot{h} = \frac{\kappa_{(4)}^2\dot{\rho}}{3\Big[2h - \frac{\epsilon}{r_{c}}(1+8\alpha
	h^2)\Big]} \, ,
\label{hderivative}
\end{equation}
thereby, $\dot{H}$  reads
\begin{equation}
\dot{H} =   \frac{\kappa_{(4)}^2h\dot{\rho}}{3H\Big[2h - \frac{\epsilon}{r_{c}}(1+8\alpha
	h^2)\Big]} + \frac{k}{a^2}\,  . \label{hderivative2}
\end{equation}
By replacing $\dot{\rho}=-3H\rho$ from Eq.~(\ref{eq23}) in the above equation, it  becomes
\begin{equation}
\dot{H} =   \frac{-\kappa_{(4)}^2\rho r_c\, ha^2+k\Big[2r_ch - \epsilon(1+8\alpha
	h^2)\Big]}{a^2\Big[2r_ch - \epsilon(1+8\alpha
	h^2)\Big]} \,  . \label{eq24}
\end{equation}
If $\alpha = 0$, i.e., a DGP model without a GB term, then the denominator of $\dot{H}$ becomes zero if $h = \pm 1/2r_c$. And this solution does not lead to any SCS. 
For each value of $\epsilon=\pm 1$, corresponding to each contracting branch $(\pm)$, the denominator of the above equation allows us to obtain two quadratic equations,
\begin{subequations}
\begin{eqnarray}
&& 8\alpha h^2+2r_ch+1=0, \quad \quad  (\epsilon=-1),\\
&&8\alpha h^2-2r_ch+1=0, \quad\quad  (\epsilon=+1).
\end{eqnarray}
\end{subequations}
Each of these equations has two real roots,
\begin{subequations}
\label{h-pm1tot}
\begin{eqnarray}
\mathfrak{h}_{\pm}^{(1)} &= - \frac{r_{c}}{8\alpha}\left(1 \mp \sqrt{1-3b}\right)  & \quad \textrm{for}~(-)~\textrm{branch}, \label{h-pm2}\quad \quad \\
\mathfrak{h}_{\pm}^{(2)} &=  \frac{r_{c}}{8\alpha}\left(1 \pm \sqrt{1-3b}\right)  & \quad \textrm{for}~(+)~\textrm{branch}. \label{h-pm1} 
\end{eqnarray}
\end{subequations}
In terms of the dimensionless functions, the equations above can be mapped, respectively, to
\begin{subequations}
	\label{h-pm-dimensionless}
	\begin{eqnarray}
	\bar{\mathfrak{h}}_{\pm}^{(1)}\, &=& \, - \frac{1}{3}\left(1 \mp \sqrt{1-3b}\right), \label{h-pm2-dimensionless}\\
	\bar{\mathfrak{h}}_{\pm}^{(2)}\, &=& \,  \frac{1}{3}\left(1 \pm \sqrt{1-3b}\right) .
	\label{h-pm1-dimensionless}
	\end{eqnarray}
\end{subequations}
In Eqs.~(\ref{h-pm-dimensionless}) the  $\pm$ signs  have nothing to do with the signs of the branches encoded in  $\epsilon$. These signs denote the two different roots of the denominator of Eq.~(\ref{eq24}) for each branch only, whereas, the solutions associated with  the negative and positive branches are now  denoted by the superscripts $(1)$ and $(2)$, respectively. Moreover, observe that $\bar{\mathfrak{h}}_{\pm}^{(2)}=- \bar{\mathfrak{h}}_{\mp}^{(1)}$, thus,  asymptotic solutions for one branch are the negative of the asymptotic solutions for the other branch.
Notice also that, for the case of parameter $b>\frac{1}{3}$, the solutions (\ref{h-pm1tot})  are not real, so they are not physical solutions. Therefore, the real solutions $\bar{\mathfrak{h}}_{\pm}^{(i)}$ (with $i=1,2$) are only possible within the range of the parameter $0<b\leq \frac{1}{3}$. 

For the $(-)$ branch, the solution $\bar{\mathfrak{h}}_{+}^{(1)}$ in Eq.~(\ref{h-pm2-dimensionless}) is associated with the energy density $\bar\rho_B$, which is always negative. Therefore, $\bar{\mathfrak{h}}_{+}^{(1)}$ is not a physically reasonable solution. Likewise, solution $\bar{\mathfrak{h}}_{-}^{(2)}$ corresponds to the energy density $\bar\rho_B<0$, which neither represents a relevant solution for the collapse scenario. Moreover, the remaining solutions are the known solutions we have found earlier, i.e. $\bar{\mathfrak{h}}_{-}^{(1)}=-\bar{\mathfrak{h}}_{+}^{(2)}=\bar{h}_A$.

The existence of the real solutions (\ref{h-pm1tot})  implies that, as the  collapse approaches the regime in which  $\bar{h}$ reaches any of the constants $\bar{\mathfrak{h}}^{(i)}_{\pm}$ above, while the values of  $\bar{H}$ and the energy density $\bar\rho$ of the brane remain finite,  the first time derivative $\dot{H}$ diverges. Therefore, as we will see in the following paragraph, the collapse would end up with an {\em abrupt} event  at a non-zero physical radius, i.e., at some $R(t_s, r)\neq0$ before the collapse reaches  the center, $R=0$, where the SFS would have been located. 
An analog abrupt event  occurs in cosmological scenarios within  phantom dark energy models; it is called a sudden singularity \cite{Barrow:2004xh}. Despite the analogy between the abrupt event appearing in our collapse model and that in cosmology, which both happen within a finite comoving time, in the present context the singularity may not be reached during a finite time coordinate of an external observer (due to formation of the horizon). Therefore, to distinguish the abrupt event occurring in our model from the one emerging in the dark energy cosmological scenarios, we term it as   sudden collapse singularity.

In order to identify which quantities $\bar{\mathfrak{h}}_\pm^{(i)}$ (with $i=1,2$ for two branches) given by (\ref{h-pm-dimensionless}) belong to the trajectories of the  dust cloud satisfying any of the solutions $\bar{h}_1, \bar{h}_2, \bar{h}_4, \bar{h}_5, \bar{h}_6$, we first set 
\begin{eqnarray}
\bar{h}_j  =  \bar{\mathfrak{h}}_\pm^{(i)} \qquad (j=1,\dots, 6),
\label{sol-s1}
\end{eqnarray}
for every physical relevant solution $\bar{h}_j$ we obtained in the previous section. If the above equation respects the  physical ranges of validity for the  parameters of the system, namely, $\eta, \theta, \vartheta$ or $\bar{\rho}_j$, then
such solutions will govern, through Eq.~(\ref{sol-s1}),  the final fate of the gravitational collapse from the formation of a SCS. Otherwise, the collapse end state will be a shell-focusing singularity. We will give a separate, more detailed explanation for each branch in the following subsections.

\subsubsection{The $(-)$ branch}

In the  following list, we further analyze  the features of this branch:
	
	\begin{enumerate}[label=\roman*)]
	    \item  
	    In the range  $0<b<\frac{1}{4}$ 
and for the initial condition $\bar{\rho}_0>\bar{\rho}_A$, the evolution of the collapse is governed by the solution $\bar{h}_1(\eta)$, given by Eq.~(\ref{EXP1}).
By setting 
$\bar{h}_1(\eta) \ = \ \bar{\mathfrak{h}}_{\pm}^{(1)}$,
we obtain $\cosh(\eta_\pm)=\pm \frac{1}{2}$. This solution implies that  the quantities (\ref{h-pm-dimensionless}) do not lie on the trajectory of the solution (\ref{EXP1}),  because the allowed range of the {\em hyperbolic cosine} term is $\cosh(\eta/3)>1$ for all $\eta>0$. In other words,  the possible values of the energy density of the brane to assure the equality (\ref{sol-s1}) read
\begin{equation}
\bar{\rho}_1(\eta_{\pm})  = \frac{2}{27}\left[1\pm \frac{1}{2}\sqrt{(1-3b)^3}\right]-\frac{b}{3}\, .
\label{rho-pm1}
\end{equation}
These  stay out of the high energy regime (with $\bar{\rho}>\bar{\rho}_0>\bar{\rho}_A$), that is, $\bar{\rho}_1(\eta_{\pm})<\bar{\rho}_A$.
This indicates that, as the collapse proceeds from its initial condition $\bar{\rho}_0>\bar{\rho}_A$ toward the center,   the brane energy density $\bar{\rho}_1(\eta)$, 
$\bar{H}_1(\eta)$ and its time derivative increase  and blow up at $R=0$. In this case, the collapse will end up with a central shell-focusing singularity. 

\item For the initial dust density $\bar{\rho_0}\leq\bar{\rho}_A$,  the solutions $\bar{h}_2(\theta)$ and $\bar{h}_4(\theta)$ represent dynamical behaviors of the collapse  within a low energy regime, where  $\theta$ belongs to the range \mbox{$0<\theta<\theta_0<\pi$}. Then, by setting $\bar{h}_2(\theta)$ into Eq.~(\ref{sol-s1}) we obtain 
$\cos\big[(\pi +\theta_\pm)/3\big] = \mp \frac{1}{2}$, or equivalently, $\theta_{+}=\pi$ and $\theta_{-}=0$.
Thereby, the possible quantities (\ref{h-pm-dimensionless})  for the $(-)$ branch that match with the solution $\bar{h}_2(\theta)$ are $\theta_{-}=0$ only; more clearly in the limit $\theta\rightarrow0$, we have $\bar{h}_2(\theta)\rightarrow\bar{\mathfrak{h}}^{(1)}_{-}~(=\bar{h}_A^\prime)$
and $\bar{\rho}_2(\theta)\rightarrow\bar{\rho}_A$.
%%%
The physical radius $R_{A}$ at which  the collapse meets the solution $\bar{\mathfrak{h}}_{-}^{(1)}$, with the energy density $\bar{\rho}_{A}$, is determined by Eq.~(\ref{sudden-radius-1}).  
Then, by using   Eq.~(\ref{Hubble0}), the finite (dimensionless) 
rate $\bar{H}_A^\prime$    associated with the solution $\mathfrak{h}_{-}^{(1)}$ is obtained as
\begin{equation}
\bar{H}_A^\prime\, =\, -\left[\left(\bar{h}_A^\prime\right)^2-\frac{\bar{k}^2}{a_0^2}\left(\frac{\bar{\rho}_{A}}{\bar{\rho}_{0}}\right)^{\frac{2}{3}}\right]^\frac{1}{2}.
\label{sudden-Hubble}
\end{equation}  
The above argument implies that, as the collapse, governed by $\bar{H}_2(\theta)$, proceeds toward the limiting regime with a finite value of  $\bar{H}_A^\prime$ and a finite energy density $\bar{\rho}_{A}$,  the time derivative $\dot{\bar{H}}_2(\theta)$ diverges. Therefore, the collapse will end up with a SCS at the limiting regime.

\item The other solution in the low energy regime, i.e., $\bar{h}_4(\theta)$,  within the given range of the parameter $\theta$, does not intersect any of the quantities (\ref{h-pm-dimensionless}). Because in this case Eq.~(\ref{sol-s1}) yields  
$\cos\left(\theta_\pm/3\right) =   \pm \frac{1}{2}$, from which we get $\theta_{+}=\pi \notin (0, \theta_0)$ and $\theta_{-}=2\pi \notin (0, \theta_0)$.
The collapse associated with  this solution, when reaching the limiting regime as $\bar{h}_4(\theta)\rightarrow\bar{h}_A$ [cf. Eq.~(\ref{limitingone})], with $\bar{h}_A\neq \mathfrak{h}_{\pm}^{(1)}$, remains regular. Afterward, it will enter the high energy domain, so that its final state will be governed by a different  dynamical evolution in this regime. Consequently, the dynamical behavior of the collapse is determined from $\bar{h}_1(\eta)$, given by Eq.~(\ref{EXP1}). This is the same solution whose evolution was discussed at the beginning of the paragraph above  Eq.~(\ref{rho-pm1}), for the initial condition $\bar{\rho}_0>\bar{\rho}_A$.
It turns out that,  if the collapsing cloud enters the high energy regime, its final state will be a SFS.

\item In the parameter range $\frac{1}{4}\leq b<\frac{1}{3}$, the physical trajectories  of the collapse are also given by Eq.~(\ref{EXP1}). Then, similar to the previous case (high energy regime), there is no solution for the equation $\bar{h}_1(\eta)  =  \bar{\mathfrak{h}}_{\pm}^{(1)}$, thus the collapse will proceed continuously toward the center without reaching any of the values $\bar{\mathfrak{h}}_{\pm}^{(1)}$. Therefore, the collapse  end state in this case is also a shell-focusing singularity.

\item For the case $b=\frac{1}{3}$, by replacing the solution $\bar{h}_5$, given by Eq.~(\ref{b=1/3case}), into Eq.~(\ref{sol-s1}) we can find the energy densities at which  quantities $\bar{\mathfrak{h}}_{+}^{(1)}=\bar{\mathfrak{h}}_{-}^{(1)}=-\frac{1}{3}$ intersect with $\bar{h}_5$ by setting $\bar{h}_5(\bar{\varrho}_{\pm})=\bar{\mathfrak{h}}_{\pm}^{(1)}$; they read $\bar{\varrho}_{+}=\bar{\varrho}_{-}=\bar{\rho}_{A}=\bar{\rho}_{B}=-\frac{1}{27}$.
This implies that none of the quantities $\bar{\mathfrak{h}}_{\pm}^{(1)}$ lie on the trajectory of the solution (\ref{b=1/3case}). Then, as the collapse proceeds, $\bar{H}_5$ and its energy density $\bar{\rho}_5$ increase until the center where they diverge. Thus, the collapse will end up with a SFS.

\item Finally, in the range of parameter $b>\frac{1}{3}$, there exist no real solutions (\ref{h-pm1tot}), thereby, no SCS would form prior to formation of the central singularity. Therefore, the final state of the collapse for this range of $b$ is  a  SFS.
\end{enumerate}

In Table~\ref{Normal-Branch},  we have summarized the fate of the collapse in terms of different initial conditions and the range of the parameter $b$. As a summary of the  above discussion, the only solution that ends up with SCS is $\bar{h}_2(\theta)$, which belongs to the range $0<b<\frac{1}{4}$. The other solutions will end up in a shell-focusing singularity. 
%%%%%%%%%%%%%%%%%%%%%%%%%%%%%%%%%%%%%%%%%%%%%%%%%%
	\begin{center}
\begin{table*}[ht!]
		\begin{tabular}{  ccccccc }
			\Xhline{2\arrayrulewidth}
			\hline\hline
			\vspace{-2mm}\\
			$b$  & \multicolumn{1}{c}{\begin{tabular}[c]{@{}c@{}} { Initial} \\   { conditions} \end{tabular}  }   & \quad $R$ \quad &~~~ \qquad   $\bar\rho$ \qquad ~~~ & ~~~ $\bar{H}, \bar{\mathcal{H}}$ ~~~  & ~~~\qquad  $\dot{\bar H}, \dot{\bar{\mathcal{H}}}$ ~~~\qquad  &  \multicolumn{1}{c}{\begin{tabular}[c]{@{}c@{}} {Nature of} \\   { singularity} \end{tabular}  } \vspace{1mm} \\ 
			\hline\\
			\multicolumn{1}{c}{\underline{\textbf{The $\boldsymbol{(-)}$ branch:}}} \vspace{3mm}\\ 
			%%-------------------------------------------------------------------------------
			$0<b<\frac{1}{4}$ & $\bar\rho_0<\bar\rho_{A}$ & $R_A$   & $\bar\rho_1(\theta)\rightarrow\bar\rho_A$  &  $\bar{H}_2\rightarrow \bar{H}_A^\prime$   &  $\dot{\bar{H}}_2\to+\infty$ &  SCS  \vspace{2mm}\\
			%%------------------------------------------------------------------------------
			& $\bar\rho_0<\bar\rho_{A}$  & $0$ & $\bar\rho_4(\theta)\rightarrow\infty$ &  $ \bar{H}_4\rightarrow-\infty$ & $\dot{\bar{H}}_4\to-\infty$ & SFS\vspace{4mm}\\
			%%------------------------------------------------------------------------------
			& $\bar\rho_0>\bar\rho_{A}$  & $0$ & $\bar{\rho}_1(\eta)\rightarrow\infty$ &  $\bar{H}_1\rightarrow-\infty$ & $\dot{\bar{H}}_1\to-\infty$ & SFS\vspace{4mm}\\
			%%-------------------------------------------------------------------------
			%%===================================================================
			$\frac{1}{4}\leq b<\frac{1}{3}$ & For all $\bar\rho_0$  & $0$ & $\bar\rho_1(\eta)\rightarrow\infty$ &  $\bar H_1\rightarrow-\infty$ & $\dot{\bar{H}}_1\to-\infty$ & SFS  \vspace{2mm}\\
			%%===================================================================
			$b=\frac{1}{3}$ &  For all $\bar\rho_0$  & $0$ & $\bar\rho_5\rightarrow \infty$ &  $\bar H_5\rightarrow -\infty$ & $\dot{\bar{H}}_5\to-\infty$ & SFS \vspace{2mm}\\
			%%--------------------------------------------------------------------
			$b>\frac{1}{3}$ & For all $\bar\rho_0$  & $0$ & $\bar{\rho}_6(\vartheta)\rightarrow\infty$ &  $\bar{H}_6\rightarrow -\infty$ & $\dot{\bar{H}}_6\to-\infty$  & SFS \vspace{2mm}\\
			%%------------------------------------------------------------------------------
			%%%%%%%%%%%%%%%%%%%%%%%%%%%%%%%%%%%%%%%%%%%%%%%%%%%%%%%%%%%%%%%%%%%%%%%
			\\
			\multicolumn{1}{c}{\underline{\textbf{The $\boldsymbol{(+)}$ branch:}}}  \vspace{3mm}\\ 
			$0<b<\frac{1}{4}$ & $\bar\rho_0<\bar\rho_{A}$ & $R_A$   & $\bar\rho(\theta)\rightarrow\bar\rho_A$  &  $\bar{\mathcal{H}}_2\rightarrow \bar{H}_A^\prime$   &  $\dot{\bar{\mathcal{H}}}_2\to-\infty$  &  SCS  \vspace{2mm}\\
			%%------------------------------------------------------------------------------
			& $\bar\rho_0<\bar\rho_{A}$  & $0$ & $\bar\rho_4(\theta)\rightarrow\infty$ &  $ \bar{\mathcal{H}}_4\rightarrow-\infty$ & $\dot{\bar{\mathcal{H}}}_4\to+\infty$ & SFS\vspace{4mm}\\
			%%------------------------------------------------------------------------------
			& $\bar\rho_0>\bar\rho_{A}$  & $0$ & $\bar{\rho}_1(\eta)\rightarrow\infty$ &  $\bar{\mathcal{H}}_1\rightarrow -\infty$ & $\dot{\bar{\mathcal{H}}}_1\to+\infty$ & SFS\vspace{4mm}\\
			%%-------------------------------------------------------------------------
			%%===================================================================
			$\frac{1}{4}\leq b<\frac{1}{3}$ & For all $\bar\rho_0$  & $0$ & $\bar\rho_1(\eta)\rightarrow \infty$ &  $\bar{\mathcal{H}}_1\rightarrow -\infty$ & $\dot{\bar{\mathcal{H}}}_1\to+\infty$ & SFS  \vspace{2mm}\\
			%%===================================================================
			$b=\frac{1}{3}$ &  For all $\bar\rho_0$  & $0$ & $\bar\rho_5\rightarrow \infty$ &  $\bar{\mathcal{H}}_5\rightarrow -\infty$ & $\dot{\bar{\mathcal{H}}}_5\to+\infty$ & SFS \vspace{2mm}\\
			%%--------------------------------------------------------------------
			$b>\frac{1}{3}$ & For all $\bar\rho_0$  & $0$ & $\bar{\rho}_6(\vartheta)\rightarrow\infty$ &  $\bar{\mathcal{H}}_6\rightarrow -\infty$ & $\dot{\bar{\mathcal{H}}}_6\to+\infty$  & SFS \vspace{2mm}\\
			%%--------------------------------------------------------------------
			\hline\hline
			\Xhline{2\arrayrulewidth}
		\end{tabular}
		\caption{Final state of the gravitational collapse on the DGP brane with the curvature term induced from the bulk, in terms of different values of the parameter $b$ and the initial condition of the collapse. There are two types of singularities: SFS  and SCS.}
		\label{Normal-Branch}
	\end{table*}
\end{center}
%%%%%%%%%%%%%%%%%%%%%%%%%%%%%%%%%%%%%%%%%%%%%%%%%%

\subsubsection{The $(+)$ branch}

In the  following,  we  complete 
our analysis for the $(+)$ branch:

\begin{enumerate}[label=\roman*)]
    \item 
    For the range  $0<b<\frac{1}{4}$ with the initial energy density $\bar{\rho}_0>\bar{\rho}_A$, by setting $\bar{\mathpzc{h}}_2(\theta)$ [cf. Table \ref{plus-Branch}] into Eq.~(\ref{sol-s1}),  we obtain two solutions:
$\cos\big[(\pi +\theta_{\pm})/3\big] =  \pm \frac{1}{2}$.
From the allowed range of the parameter $\theta$, we observe that the only quantity that lies on the path of the solution $\bar{\mathpzc{h}}_2(\theta)$, in parameter space of  $\theta$,  is $\bar{\mathfrak{h}}_{+}^{(2)}$. Note that $\bar{\mathfrak{h}}_{+}^{(2)}=-\bar{h}_A^\prime$, being the value of $\bar{h}$ at the limiting regime. 
The physical radius $R_{A}$ at which  $\bar{\mathpzc{h}}_2\rightarrow-\bar{h}_A^\prime$ and, subsequently,  $\bar{\rho}_2(\theta)\rightarrow\bar{\rho}_{A}$, is determined by Eq.~(\ref{sudden-radius-1}).  
Then, by using   Eq.~(\ref{Hubble0}), the finite (dimensionless) $\bar{H}_A^\prime$    associated with the solution $\mathfrak{h}_{+}^{(2)}$ is obtained as
\begin{equation}
\bar{H}_A^\prime\, =\, -\left[\left(-\bar{h}_A^\prime\right)^2-\frac{\bar{k}^2}{a_0^2}\left(\frac{\bar{\rho}_{A}}{\bar{\rho}_{0}}\right)^{\frac{2}{3}}\right]^\frac{1}{2}.
\label{sudden-Hubble2}
\end{equation}  
This implies that, as the collapse tends to the limiting regime, $\bar{\mathcal{H}}_2(\theta)$ and the energy density $\bar{\rho}_2(\theta)$ of the brane remain \emph{finite}, whereas the time derivative $\dot{\bar{\mathcal{H}}}_2(\theta)$, diverges. Therefore, the final state of the collapse in this case will be a SCS.

\item On the other hand, for the solution $\bar{\mathpzc{h}}_4(\theta)$, Eq.~(\ref{sol-s1}) yields $\cos(\theta_{\pm}/3) =  \mp 1/2$. None of the parameter $\theta$, i.e. $\theta_{\pm}=\pi, 2\pi$ lie on the trajectory of $\bar{\mathpzc{h}}_4$. Then, as the collapse proceeds, it passes through the limiting regime and enters the high energy regime. Hereafter, the collapse fate will be determined by its evolution due to  the solution $\bar{\mathpzc{h}}_1(\eta)$. In the high energy regime, by setting $\bar{\mathpzc{h}}_1(\eta)$ in Eq.~(\ref{sol-s1}), we obtain
$\cosh(\eta_{\pm}/3) =  \mp 1/2$. This solution is not physically relevant  because $\eta>0$, so $\cosh(\eta/3)>1$. Thus, none of the quantities $\bar{\mathfrak{h}}_{\pm}^{(2)}$ can be reached as the collapse evolves. Therefore, the evolution will continue until the dust cloud reaches the center, where a shell-focusing singularity will form.

\item For $\frac{1}{4}\leq b<\frac{1}{3}$ the solution is given by $\bar{\mathpzc{h}}_1(\eta)$, {\em i.e.}, the same solution as that of the high energy regime. Therefore, none of the quantities (\ref{h-pm-dimensionless}) lie on the dust cloud trajectory, and likewise the collapse final state will be a shell-focusing singularity.

\item For  $b=\frac{1}{3}$, the energy density associated with the solution $\bar{\mathpzc{h}}_5$ reads $\bar{\rho}_{A}=\bar{\rho}_{B}=-\frac{1}{27}<0$.  This is equal to the energy density associated with  Eqs.~(\ref{h-pm-dimensionless}) which are not physically reasonable  for the  brane's energy densities. 
This implies that the quantities $\bar{\mathfrak{h}}_{\pm}^{(2)}$  cannot be endorsed as solutions for the collapse end state. Thus, the brane will proceed with  increasing  $\bar{\mathcal{H}}_5$ and  $\bar{\rho}_5$ until it reaches the center. So, the collapse end state will be a shell-focusing singularity.

\item Finally, for the range $b>\frac{1}{3}$,  the  solutions (\ref{h-pm1tot}) are not real,   so this does not correspond to a physically relevant solution. Thus, none of solutions (\ref{h-pm1tot}) are viable for the collapse end state. Therefore, as discussed in Sec.~\ref{D} (see also Table \ref{plus-Branch}), the collapse process will terminate with a shell-focusing singularity.

\end{enumerate}

Here, similar to the $(-)$ branch, the only solution that ends up with SCS is $\bar{h}_2(\theta)$, which belongs to the range $0<b<\frac{1}{4}$,  and the other solutions will end up in a shell-focusing singularity. 
In the rest of the paper, we will study the physical implications of these solutions. We will answer the following questions: Do these solutions satisfy energy conditions? How do they match to the exterior space time? By answering the latter, we will be able to distinguish the solutions that will be hidden behind an event horizon from those that will be naked singularities.

\begin{figure*}
	\begin{center}
		\includegraphics[scale=0.5]{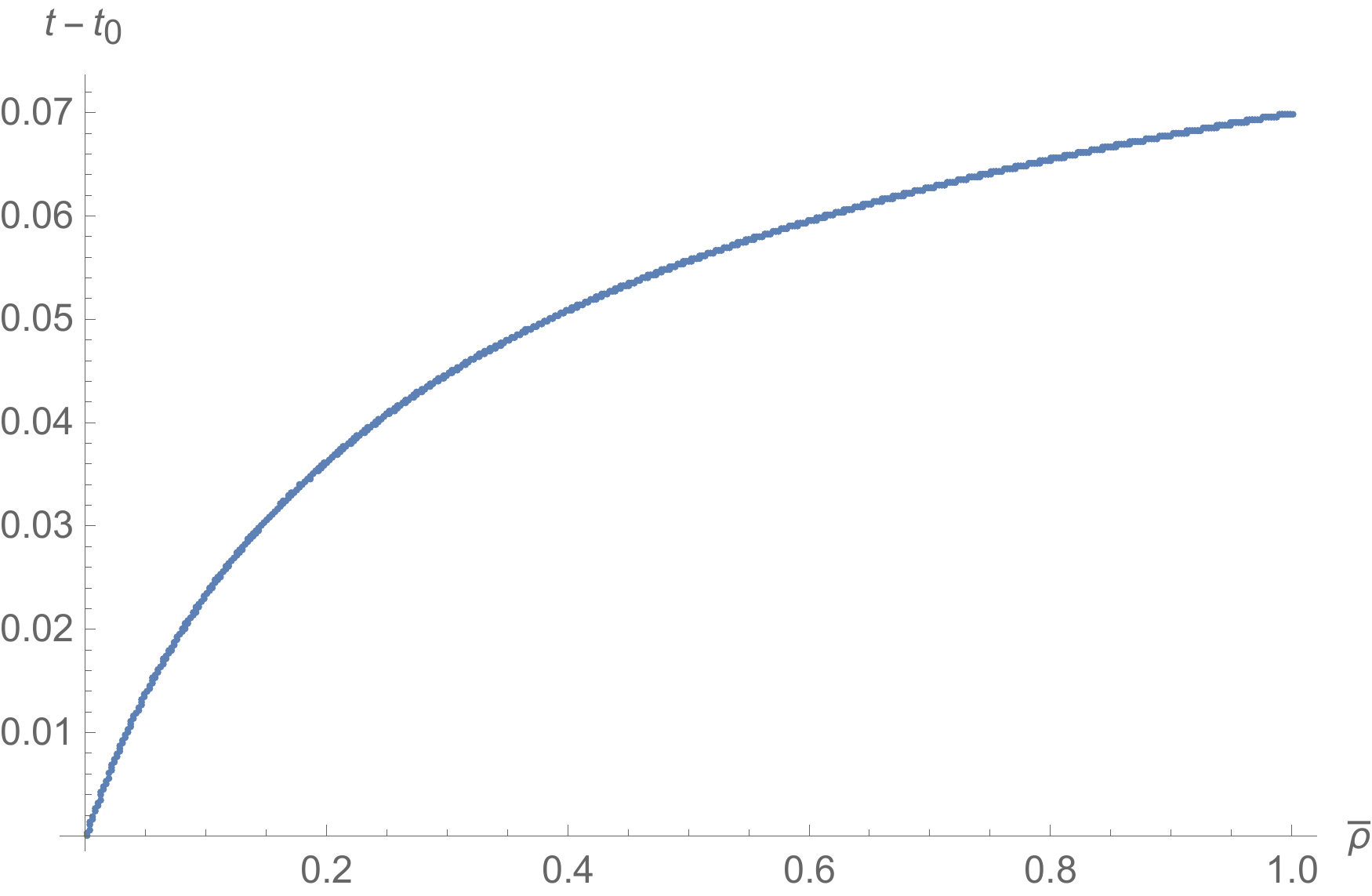}
		\caption{The time $t_{\rm sing}-t_{0}$, for the sudden singularity formation, is plotted in terms of the density $\bar{\rho}$.}\label{time}
	\end{center}
\end{figure*}

Since we are interested in the fate of the singularity, we need to establish how much time  it takes for the sudden singularity to occur. This can be obtained  by integrating the energy conservation equation with respect to the density $\rho$. Therefore, the time  $t_{\rm sing}-t_{0}$ required for the singularity occurrence reads, 
\begin{equation}
	t_{\rm sing} - t_0 = -\frac{br_c}{3}\int_{\bar{\rho}_0}^{\bar{\rho}_{A}} \frac{d\bar{\rho}}{\bar{\rho}\, \bar{H}(\bar{\rho})}\, ,
	\label{time-sing}
\end{equation}
where $t_{0} $ and $t_{\rm sing}$ are respectively, the initial time  and the time at which the singularity forms.
Figure \ref{time} depicts the time $t_{\rm sing}-t_{0}$ as a function of the density $\bar{\rho}$ for the solution (\ref{lowenergy1}). It is clear that the time interval (\ref{time-sing}) during which a SCS happens is finite. This is because both parameters   $\bar{\rho}$ and $\bar{H}(\bar{\rho})$ are bounded and nonvanishing.

%%%%%%%%%%%%%%%%%%%%%%%%%%%%%%%%%%%%%%%%%%%
\subsection{Energy conditions}

So far we have only obtained solutions to  Eq.~(\ref{modifiedfriedmann}). In this section, we examine their compatibility with a set of energy conditions \cite{Hawking:1973uf}. The ``weak energy condition''  is  needed to be satisfied by  standard matter.  It guarantees, for example, the positiveness of the energy density for any local timelike observer. 
We need to define an effective energy density $\rho_{\rm eff}$ for the collapse to follow the energy conditions discussion according to their well-known form in GR.

Let us rewrite the modified Friedmann equation (\ref{modifiedfriedmann}) in an effective relativistic form:
\begin{eqnarray}
H^2 &=& \frac{\kappa_{(4)}^2}{3}\rho_{\rm eff} -\frac{k}{a^2}\, ,
\label{modifiedfriedmann-2}
\end{eqnarray}
where, we have introduced an effective energy density $\rho_{\rm eff}$ on the brane as
\begin{eqnarray}
\rho_{\rm eff} &:=& \rho + 
\frac{6\epsilon}{\kappa_{(5)}^2}\left(1+\frac83\alpha h^2\right)h \, .
\label{energydensity-1}
\end{eqnarray}
We introduce an effective pressure by defining  an  effective equation of state $w_{\rm eff}$ for a perfect fluid with an effective energy density $\rho_{\rm eff}$,  as $p_{\rm eff}:=w_{\rm eff}\, \rho_{\rm eff}$. Thereby,  an effective conservation equation can be written as
\begin{equation}
\dot{\rho}_{\rm eff}+3H(1+w_{\rm eff})\rho_{\rm eff}=0.
\label{eff-conservation}
\end{equation}
Then, using Eq.~(\ref{energydensity-1}) we obtain
\begin{eqnarray}
\dot{\rho}_{\rm eff} &:=& \dot{\rho} + 
\frac{6\epsilon}{\kappa_{(5)}^2}\left(1+8\alpha h^2\right)\dot{h} \, ,
\label{energydensity-der}
\end{eqnarray}
where $\dot{h}$ is given by Eq.~(\ref{hderivative}). From Eqs.~(\ref{eff-conservation}) and (\ref{energydensity-der}) the effective pressure can be written as
\begin{eqnarray}
p_{\rm eff} &=& -\rho_{\rm eff} - \frac{\dot{\rho}}{3H} - 
\frac{2\epsilon}{\kappa_{(5)}^2H}\left(1+8\alpha h^2\right)\dot{h} \nonumber \\
&=& -\rho_{\rm eff} + 
\frac{2r_ch\rho}{2r_ch - \epsilon(1+8\alpha
	h^2)}  \, ,
\end{eqnarray}
where in the second term, we have used the conservation equation (\ref{eq23}) for the dust fluid.

The effective energy-momentum tensor should  satisfy the energy conditions (ECs)
\begin{subequations}
	\label{EC-orig}
	\begin{eqnarray}
	\textrm{EC.1}: \quad \quad&& \rho_{\rm eff}\ =\
	\rho + 
	\frac{6\epsilon}{\kappa_{(5)}^2}\left(1+\frac83\alpha h^2\right)h\geq 0, \label{EC1-orig} \\
	\textrm{EC.2}: \quad \quad&& \rho_{\rm eff}+p_{\rm eff} \  =\ 
	\frac{2r_ch\rho}{2r_ch - \epsilon(1+8\alpha
		h^2)} \geq 0 \, . \label{EC2-orig}
	\end{eqnarray}
\end{subequations}
The above inequality  can be rewritten in terms of dimensionless parameters as
\begin{subequations}
	\label{EC} 
	\begin{eqnarray}
	&& 
	\textrm{EC.1}: \quad \quad \quad \quad \bar\rho + \epsilon 
	\left(b+  \bar{h}^2\right)\bar{h}\geq 0 , \label{EC1} \\
	&& 
	\textrm{EC.2}: \quad \quad \quad \quad \frac{2\bar{h} \bar\rho}{2 \bar{h} - 
		\epsilon(b+ 3\bar{h}^2)} \geq 0 \, .\label{EC2} 
	\end{eqnarray}
\end{subequations}
The first energy condition (EC.1) above can be written for each branch, respectively, as
\begin{eqnarray}
&& \bar{h}^3 + b\bar{h} - \bar\rho \leq 0  \quad \quad  \textrm{for the $(-)$ branch,}\\
&& \bar{h}^3 + b\bar{h} + \bar\rho \geq 0  \quad \quad  \textrm{for the $(+)$ branch}. 
\end{eqnarray}
From Eq.~(\ref{Friedmannnb}) we observe  that ${\bar h}^3+b\bar h-\bar\rho=-{\bar h}^2<0$, thus, the EC.1 for the $(-)$ branch ({\em i.e.,} the first equation above) is satisfied.
Likewise, Eq.~(\ref{Friedmannnb2})  indicates that ${\bar h}^3+b\bar h+\bar\rho={\bar h}^2 > 0$, so
the second inequality above [i.e., EC.1 for the $(+)$ branch] is also  satisfied. Therefore, the EC.1 (\ref{EC1}) is  respected in the herein effective theory for both branches.

The second energy condition (EC.2), given by (\ref{EC2}), can be written equivalently as
\begin{subequations}
\begin{eqnarray}
\textrm{EC.2a}\colon \quad   \quad
\bar{h}>0 \quad  	\land \quad  2 \bar{h} - 
\epsilon(b+ 3\bar{h}^2)>0  ,   \quad \quad  \quad
\label{1condition}
\end{eqnarray}
or
\begin{eqnarray}
\textrm{EC.2b} \colon \quad    \quad
\bar{h}<0 \quad 	\land \quad  2 \bar{h} - 
\epsilon(b+ 3\bar{h}^2)<0.   \quad \quad \quad
\label{2condition}
\end{eqnarray}
\end{subequations}
Depending on the sign of $\epsilon$ we can distinguish two different cases as we will analyze in the following.

\subsubsection{The $(-)$ branch}	

%\noindent {\bf The $(-)$ branch.}
In this case, one of the conditions EC.2a or EC.2b should be satisfied. 
\begin{enumerate}[label=\alph*)]
	\item  If the  condition EC.2a, given by Eq.~(\ref{1condition}), holds, we would have
	\begin{equation}
	\quad \left\{\bar{h}>0\right\} \ \land \ \left\{\bar{h}<\bar{\mathfrak{h}}_{-}^{(1)} \ \lor \  \bar{h}>\bar{\mathfrak{h}}_{+}^{(1)}\right\}, 
	\label{Ec1-a} \nonumber
	\end{equation}
	where both quantities $\bar{\mathfrak{h}}_{\pm}^{(1)}$ are  negative and defined in Eq.~(\ref{h-pm-dimensionless}). The intersection of the above sets reads $\bar{h}>0$, which means that 
	the range of quantities  $\bar{h}$ that holds EC.2a is $\bar{h}>0$. This condition results in a class of physically relevant solutions,
	\begin{eqnarray}
	\quad \textrm{EC.2a}  \quad    \Leftrightarrow \quad  \left\{\bar{h}_1(\eta),\ \bar{h}_4(\theta),\ \bar{h}_5,\  \bar{h}_6(\vartheta)\right\} ,\quad  \label{EC2b-}
	\end{eqnarray}
	for the herein DGP-GB model of dust collapse (cf. Sec.~\ref{model}). In this case, $\bar{h}(t)$ is positive, having a lower bound at $\bar{k}/a(t)$ at any time $t$. But, it has no upper limit, so it can diverge at some time in the future.
	%%%%%%%%%%%%%%%%%%%%%%%%%%%%%%%%%%%%%%%%%%%%%%%%%%%%%%%%%%%%%%%%%%%%%%%%
	\item The  condition EC.2b in Eq.~(\ref{2condition}) implies that
	\begin{eqnarray}
	\quad \left\{\bar{h}<0\right\} \ \land  \  \left\{\bar{\mathfrak{h}}_{-}^{(1)}<\bar{h}<\bar{\mathfrak{h}}_{+}^{(1)}\right\}. \nonumber
	\label{Ec2-a}
	\end{eqnarray}
	Since $\bar{\mathfrak{h}}_{-}^{(1)},\, \bar{\mathfrak{h}}_{+}^{(1)}<0$,  the intersection of two ranges above will be $\bar{\mathfrak{h}}_{-}^{(1)}<\bar{h}<\bar{\mathfrak{h}}_{+}^{(1)}$.
	In this range for $\bar{h}$, there exists only one physically reasonable solution,
	\begin{equation}
	\textrm{EC.2b}  \quad    \Leftrightarrow \quad  \bar{h}_2(\theta)\, . \label{EC2b-2}
	\end{equation}
\end{enumerate}
The  sets of solutions (\ref{EC2b-}) and (\ref{EC2b-2}) display two separate scenarios: The first one (cf. EC.2a) represents a  collapse  starting  from an initial configuration at some values $\bar{h}_0>0$ and evolving by increasing $\bar{h}_2(\theta)$ toward the $R=0$, at which $\bar{H}$, $\bar{\rho}$, and $\dot{\bar{H}}$ diverge. Therefore, the collapse end state in this case is a  SFS. 
%%%%	
The second scenario (cf. EC.2b) performs a collapse starting  from an initial state with  $\bar{\mathfrak{h}}_{-}^{(1)}\leq \bar{h}_0< \bar{\mathfrak{h}}_{+}^{(1)}$ and evolves toward $\bar{\mathfrak{h}}_{-}^{(1)}$ by  (negatively) decreasing function $\bar{h}(\theta)$. When it reaches $\bar{\mathfrak{h}}_{-}^{(1)}~ (=\bar{h}_A)$, the brane hits a SCS.

\subsubsection{The $(+)$ branch}

%\noindent {\bf The $\boldsymbol{(+)}$ branch.}
For this  branch, we  also have two cases, depending on if either of the conditions EC.2a or EC.2b is satisfied.
\begin{enumerate}[label=\alph*)]
	\item Following the  condition EC.2a,  two inequalities yield the condition on $\bar{h}$ as
	\begin{eqnarray}
	   \left\{\bar{h}>0\right\} \  	\land \ \left\{\bar{\mathfrak{h}}_{+}^{(2)} < \bar{h}< \bar{\mathfrak{h}}_{-}^{(2)}\right\},       \nonumber
	\end{eqnarray}
	where $\bar{\mathfrak{h}}_{\pm}^{(2)}$ are both positive numbers  given by Eq.~(\ref{h-pm-dimensionless}).
	The intersection of two ranges above gives a physical range for the evolution of  $\bar{h}$ as
	\begin{eqnarray}
	\bar{\mathfrak{h}}_{+}^{(2)}<\bar{h}<\bar{\mathfrak{h}}_{-}^{(2)}\, . \quad 
	\label{EC2a+}	
	\end{eqnarray} 
	This  positive range of $\bar{h}$ contains only one solution (cf. Table \ref{plus-Branch}):
	\begin{eqnarray}
	 \textrm{EC.2a}  \quad  \Leftrightarrow   \quad \bar{\mathpzc{h}}_2(\theta).
	\end{eqnarray}    
	%%%%%%%%%%%%%%%%%%%%%%%%%%%%%%%%%%%%%%%%%%%%%%%%%%%%%%%%%%%%%%%%%%%
	\item Likewise, for the  condition EC.2b, we get
	\begin{equation}
	 \left\{\bar{h}<0\right\} \ \land \ \left\{\bar{h}<\bar{\mathfrak{h}}_{+}^{(2)} \ \lor \  \bar{h}>\bar{\mathfrak{h}}_{-}^{(2)}\right\}.
	\label{Ec1+a} \nonumber
	\end{equation}
	The intersection of the above sets provides a  range of changes for $\bar{h}$ as 
	\begin{eqnarray}
	\bar{h}<0\, . \quad  
	\end{eqnarray}
	This condition implies that the physically reasonable solutions for the $(+)$ branch in this case is given by the solutions (cf. Table \ref{plus-Branch}):
	\begin{eqnarray}
	\textrm{EC.2b}  \quad  \Leftrightarrow   \quad \left\{\bar{\mathpzc{h}}_{1}(\eta),\ \bar{\mathpzc{h}}_{4}(\theta),\ \bar{\mathpzc{h}}_{5},\ \bar{\mathpzc{h}}_{6}(\vartheta)\right\}.
	\end{eqnarray}
\end{enumerate}
The above argument describes two different scenarios for the herein collapse on the $(+)$ branch: The case EC.2a  describes a collapse  starting  at an initial configuration with $\bar{h}_0$ holding the range $\bar{\mathfrak{h}}_{+}^{(2)}< \bar{h}_0< \bar{\mathfrak{h}}_{-}^{(2)}$ and then evolves by an incremental function $\bar{h}$ until  it reaches the upper limit $\bar{\mathfrak{h}}_{-}^{(2)}=-\bar{h}_A^\prime$. In this limit, the brane ends up with  a SCS. The second scenario, EC.2b, corresponds to a collapse that began from an initial state with  $\bar{h}_0< 0$ and evolves (by   $\bar{h}$ decreasing negatively) toward the center. When it reaches $R=0$, the brane ends with a shell-focusing singularity.

As a consequence of the analysis we have presented in this section, we have found differences with respect to the scenarios  provided by GR; therein,   the energy densities 
always blueshift and eventually diverge, with the central shell-focusing (naked or black hole) singularities emerging. 
Instead, herewith our investigation within the  DGP-GB modifications to the dynamics of the collapse, we have found a range for the parameter $b$ that leads to a different destiny for the collapse outcome. In particular, by choosing a convenient initial condition as $\rho_0 < \rho_A$, $\dot{h}_0<0$ and for the range of parameter $0<b< \frac{1}{4}$, the collapse of a dust fluid on the DGP brane would terminate with a SCS abrupt event, rather different
from the  mere formation of a central singularity. In the next section, we will investigate the circumstances for the formation of black hole horizons.

%%%%%%%%%%%%%%%%%%%%%%%%%%%%%%%%%%%%%%%%%%%%%%%%%%%%%%%%%%%%%%%%%%%%%%%
\section{Status of the black hole horizon}
\label{exterior}

In this section, we will present the matching conditions at the boundary of the collapsing object on the brane with a convenient exterior geometry. We then apply such conditions to the particular solutions we retrieved for the herein DGP-GB collapse.

%\subsection{Matching conditions and generic exterior function}

The modified Friedmann Eq.~(\ref{modifiedfriedmann-2}) can be written  as
\begin{eqnarray}
\dot{R}^2 =    f_0(R) +\epsilon f(R)  , 
\label{modifiedfriedmann-3}
\end{eqnarray}
where,   two functions $f_0(R)$ and $f(R)$  were introduced as
\begin{align}
	f_0(R)   &:= \frac{2Gm_0}{R} + \mathcal{E} , \quad \quad
	\\
	f(R) &:= \frac{2Gm(R)}{R} \,  =\,  \left(\bar{h}^3+b\bar{h}\right) \frac{R^2}{b^2r_c^2}\, ,
	\label{f_1}
\end{align}
in terms of the physical radius $R$, where  $\mathcal{E}\equiv-k \mathbf{r}_{\rm b}^2>-1$, and $m_0 =(4\pi \mathbf{r}_{\rm b}^3a_0^3/3)\rho_0$ is the initial mass of the star. 
The first term in Eq.~(\ref{f_1}) stands for the pure  relativistic effects, the second term stands for the DGP-GB modifications, and the term $\mathcal{E}$ is the energy per unit mass.

Since the OS collapse in GR is matched to a Schwarzschild exterior, our objective is to find a general {\em static} exterior geometry for our herein modified OS model.
Let us then match the interior modified OS spacetime at the boundary $\Sigma$, to a convenient general, spherically symmetric static  exterior metric of the form \cite{Bruni:2001fd}
\begin{equation}
ds^2_{\rm ext} = -A^2(R)F(R)d\tau^2  + F^{-1}(R)dR^2 + R^2d\Omega^2\, ,
\end{equation}
for some unknown function $A(R)$ and $F(R)$,
\begin{equation}
F(R) := 1-\frac{2GM(R)}{R}\, .
\label{redshift1}
\end{equation}
These two functions should be determined by employing a convenient set of matching conditions. 
Thus, following the approach presented in \cite{Bruni:2001fd}, we get
\begin{eqnarray}
\dot{R}^2 &=&    -F(R) + \tilde{\mathcal{E}} , \quad \quad 
\label{modifiedfriedmann-4}
\end{eqnarray}
where $\tilde{\mathcal{E}}$ is a constant. Contrasting Eqs.~(\ref{modifiedfriedmann-3}) and (\ref{modifiedfriedmann-4}), we obtain
\begin{eqnarray}
F(R)  \, =\,  \tilde{\mathcal{E}} -  f_0(R) -\epsilon f(R)  . \quad \quad 
\label{modifiedfriedmann-5}
\end{eqnarray}
Moreover, by choosing $\tilde{\mathcal{E}}=\mathcal{E}-1$, without loss of generality, we get
\begin{eqnarray}
F(R)  \, =\,  1 -  \frac{2Gm_0}{R} -\epsilon  \frac{2Gm(R)}{R}\, .
\label{modifiedfriedmann-6}
\end{eqnarray}
Using this, we can introduce a generalized mass $M(R)$ in terms of the Schwarzschild mass $m_0$ and the modifications induced by the DGP-GB effects as
\begin{eqnarray}
M(R) &=&  m_0 +\epsilon\, m(R).
%\left[L(R) +  P(R)\right]. 
\label{mass-eff}
\end{eqnarray}
This relation states that, in the presence of the DGP-GB modifications to GR, provided by the correction term  $m(R)=(R/2G)f(R)\neq 0$,  the exterior solution deviates from the standard Schwarzschild geometry. However, for the particular case of $\alpha=0$, and in the limit of larger scales  (when the distances are much larger than $r_c$), the action (\ref{action}) reduces to the four-dimensional Einstein-Hilbert  action and the  Schwarzschild solution is restored for the exterior.

In what follows, we will analyze the possible exterior geometries and the status of horizon formation at the dust boundary in terms of the the  interior collapse solutions we have derived in Sec.~\ref{model}). 
We start by assuming that the collapse is initially untrapped, that is,
the  parameters $m_0, b, r_c$, and $R_0$ are such that   $F(R)$ is initially  positive,
\begin{equation}
F(R_0) = 1 - \frac{2Gm_0}{R_0} - \epsilon f(R_0)>0\, .
\end{equation}
Then, as the collapse evolves, depending on the induced-gravity effects, $F(R)$ may or may not become zero or negative. The roots of the equation $F(R_H) = 0$ provides the location of the horizon $R_H$ of the final black hole. If such a root does not exist, then no black hole would form.

\begin{figure*}
	\begin{center}
		\includegraphics[scale=0.7]{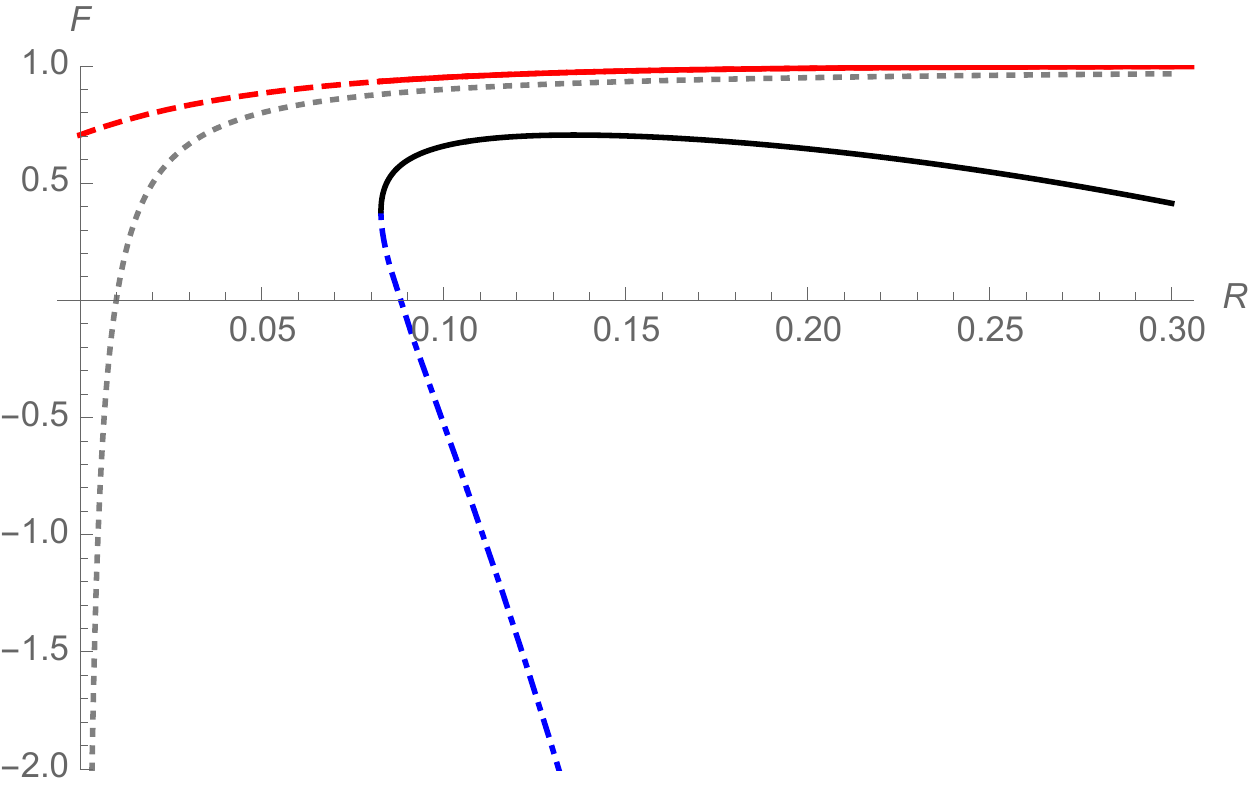}  \quad 
		\includegraphics[scale=0.7]{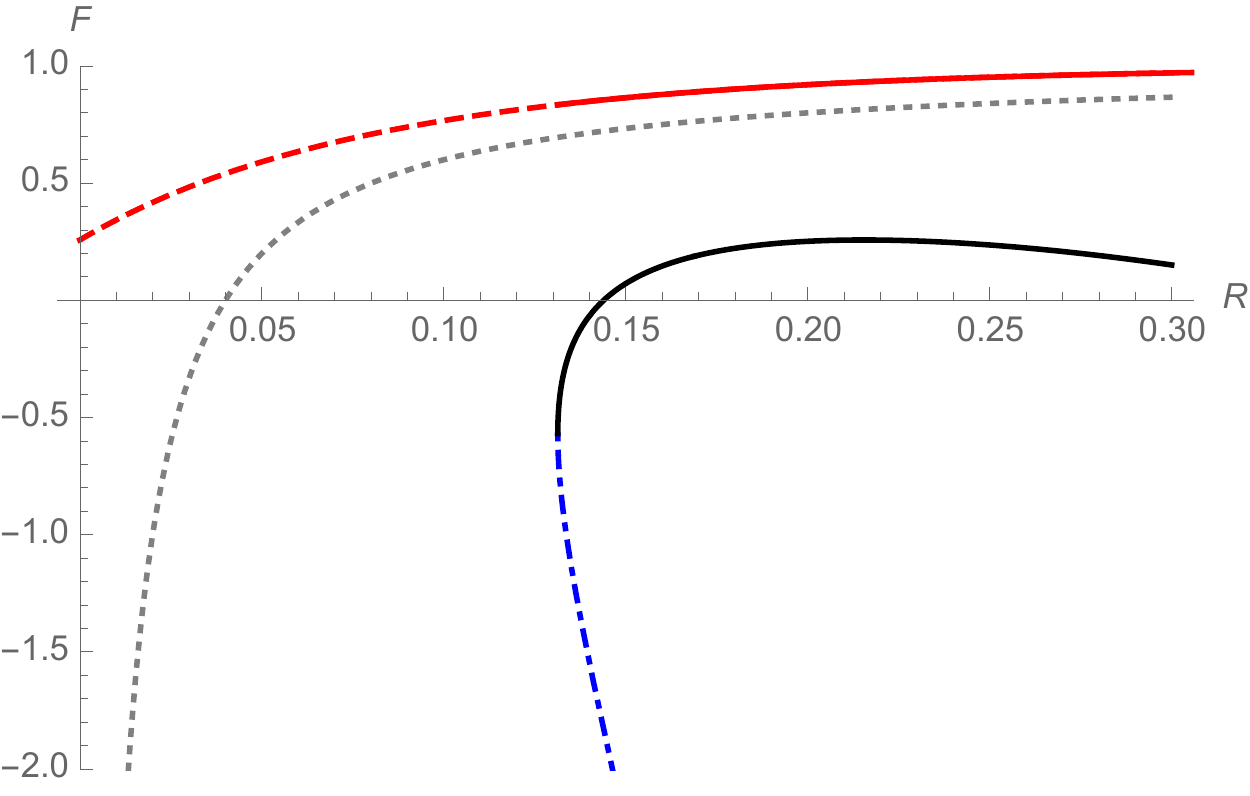}  \quad 
		\includegraphics[scale=0.7]{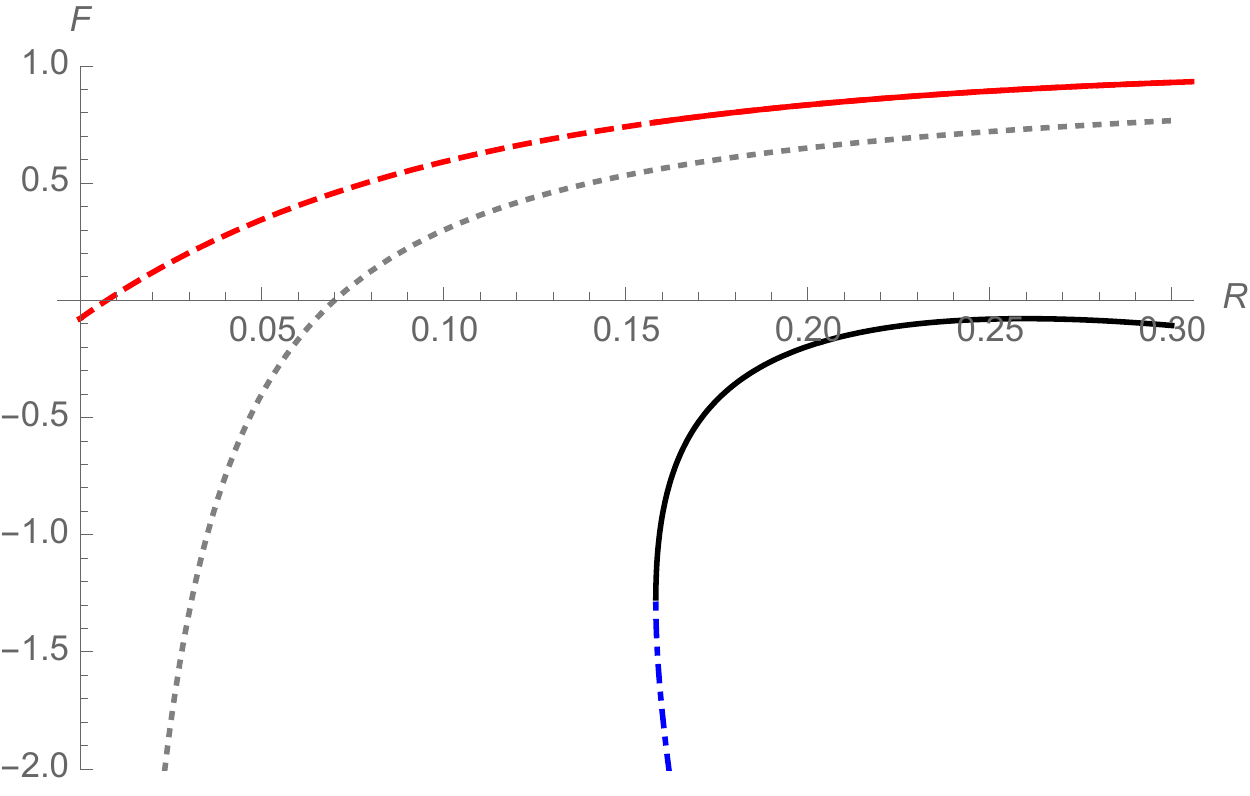} 
		\caption{The evolution of the exterior functions associated with the interior solutions  for $\epsilon = -1 $ are depicted. The black  and the red curves represent, respectively, exterior functions associated with the solutions  $\bar{h}_2$ (leading to a SCS) and  $\bar{h}_4$ (containing a SFS).  The dashed blue and red curves depict the solutions $\bar{h}_3$ (physically irrelevant) and $\bar{h}_1$ (leading to SFS), respectively.  Finally,   the dotted gray curves are associated with the standard relativistic Schwarzschild geometry with $f_0(R)=1-2Gm_0/R$.
			The collapse parameters are chosen as $b=1/8$, $G=1$, $r_c=0.5$, and the initial values of the  star mass are $m_{0} = 0.005$ (upper), $m_{0} = 0.02$ (middle), and $m_{0} = 0.035$ (lower).}\label{Exterior-1}
	\end{center}
\end{figure*}

\begin{figure*}
	\begin{center}
		\includegraphics[scale=0.7]{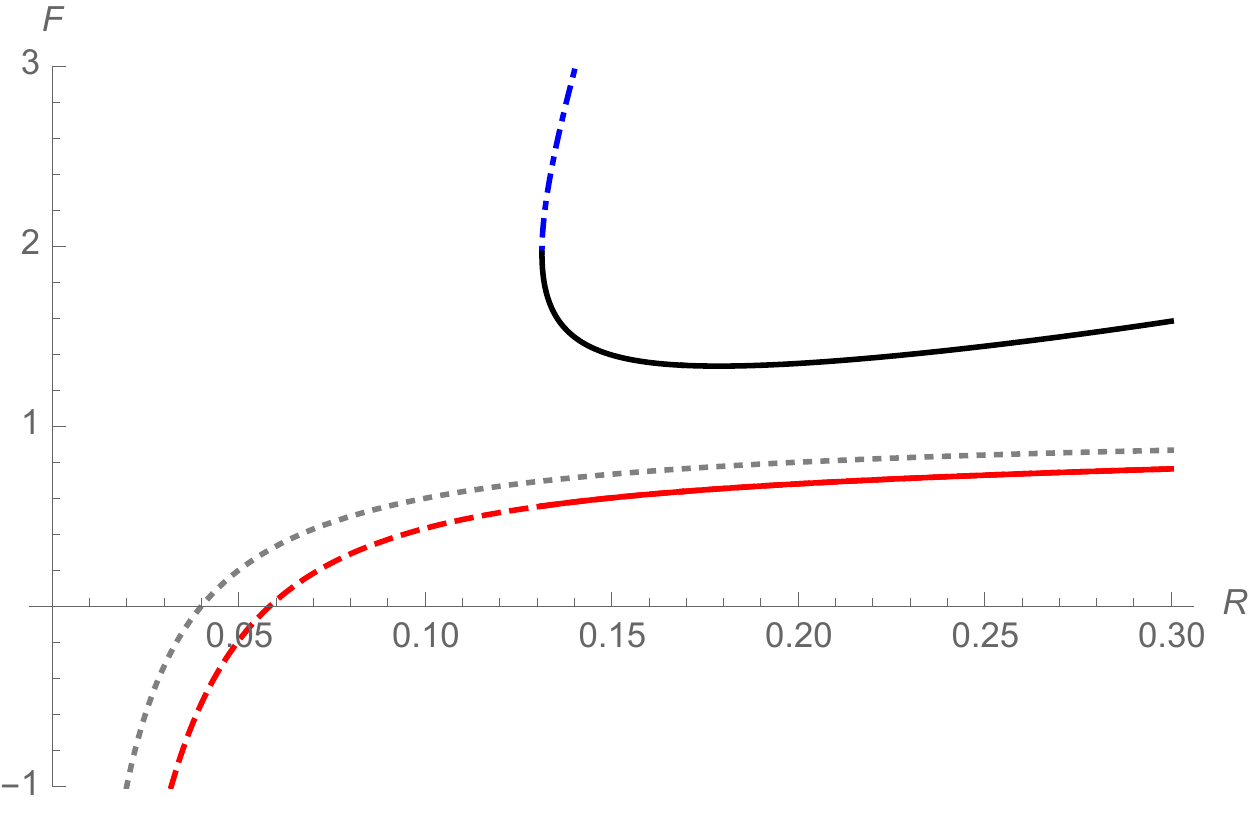} 
		\caption{The evolution of the exterior functions associated with the interior solutions  for $\epsilon = +1 $ are depicted. The black  curve represents the exterior function associated with the solution  $\bar{\mathpzc{h}}_2$ (containing a SCS).  The dashed  and solid red curves depict the solutions $\bar{\mathpzc{h}}_1$ and $\bar{\mathpzc{h}}_4$ (containing SFSs), respectively.  Finally,   the dotted gray curve is associated with the standard relativistic Schwarzschild geometry with $f_0(R)=1-2Gm_0/R$.
			The collapse parameters are chosen as $b=1/8$, $G=1$, $r_c=0.5$, and  $m_{0} = 0.02$.}\label{Exterior-2}
	\end{center}
\end{figure*}

Figures \ref{Exterior-1} and \ref{Exterior-2} present  numerical behaviors of the exterior function $F(R)$ for the physically relevant solutions we  obtained in the Sec.~\ref{model} for the two branches, $\epsilon=\pm 1$. Both diagrams were plotted for the range of parameter $0<b<\frac{1}{4}$. In  Fig.~\ref{Exterior-1}  the black  and  red curves depict  the exterior functions associated with the two solutions $\bar{h}_2$ and $\bar{h}_4$, respectively. The dashed blue and red curves represent the exterior function corresponding to the solutions\footnote{The solution $\bar{h}_3$, as mentioned earlier, is not physically reasonable for our model because it does not illustrate a low energy regime.} $\bar{h}_3$ and $\bar{h}_1$. Finally, the dashed gray curves depict the standard relativistic exterior function representing  a Schwarzschild black hole geometry with  constant mass $m_0$. 
On those figures, the points where the solid curves end and the dashed curves begin are the radii at the limiting regime $R_A$.

As it is shown in Fig.~\ref{Exterior-1},  for a set of initial parameters, there exist solutions in which the horizontal axis will not be crossed by  $F(R)$, which indicates that {\em no} horizon would  form in the collapse process and the final state of the collapse will be a {\em naked} SFS. To be more concrete, in  Fig.~\ref{Exterior-1} ($\epsilon = -1$ branch), the red  solid curve represents the exterior function of a collapse that begins from a low energy regime [cf. the solution $\bar{h}_4$ in Eq.~(\ref{lowenergy3})] and evolves toward the limiting regime at $R_A$. 
Subsequently, it enters the high energy regime [governed by the solution $\bar{h}_1$, Eq.~(\ref{EXP1})] and proceeds toward the center. Depending on the choice of initial parameters, the exterior function may intersect the horizontal axis. If it reaches the horizontal axis, the final state for the exterior geometry will be a black hole. Likewise, for the  solid black curve, there is a range of initial parameters for which, as the collapse proceeds from its initial condition up to the limiting regime at the radius $R_A$, no intersection between exterior function and the horizontal axis will happen; hence the exterior geometry will represent  a naked  sudden singularity. For other choices of the parameters, an apparent horizon would form and the final sudden singularity would be hidden by a black hole horizon.

Figure \ref{Exterior-2}  depicts how the exterior geometry of the $(+)$ branch behaves differently. The exterior function associated with the interior solution $\bar{\mathpzc{h}}_{2}$ (which leads to a SCS)  does not have an event horizon regardless of the initial mass $m_{0}$ value, thus the SCS is naked. On the other hand, the solutions $\bar{\mathpzc{h}}_{1}$ and $\bar{\mathpzc{h}}_{4}$, which result in SFSs, are always hidden inside the event horizon of a black hole.

%%%%%%%%%%%%%%%%%%%%%%%%%%%%%%%%%%%%%%
\section{Conclusions and Outlook}
\label{conclusion}

In this paper, we investigated a higher-dimensional brane world extension of the general relativistic OS model for gravitational collapse.
More precisely, we considered a DGP brane world model for gravitational collapse, where a GB term was present on the bulk. The brane was filled with a homogeneous dust fluid as matter content, whereas its geometry was governed by a closed FLRW metric.

From extracting the evolution equation of the brane, we subsequently provided a class of solutions relevant to describe a contracting brane. This conveys the gravitational collapse of a spherically symmetric object toward the central singularity.

We also obtained a particular solution in which, as the brane evolves toward a minimum nonzero radius, the energy density and $\bar H$ remain finite; the first time derivative of $\bar H$ diverges, though. Therefore, the brane reaches a particular singularity before it could reach the central singularity for which the energy density and the curvature of the brane become infinite. 
Such singularity is known as sudden singularity which would occur in a finite cosmic time in the future. Nevertheless, given that the singularity is reached in an infinite time from the point of view of an external observer, we will call it a sudden collapse singularity.

By employing a convenient matching condition at the boundary of the collapsing dust, we also provided an exterior solution for the collapse final state. By doing so, we were able to examine the role of the GB sector on the process and the outcome of the collapse.  We have shown that  for the    $(+)$ branch, there exists a solution where the SCS happens before the formation of an event horizon. This occurs regardless of the initial conditions of the system. This is the direct consequence of the GB term which modifies the system in the high energy regime. The other shell-focusing solutions will have an event horizon no matter what initial settings they started with.  On the other hand, in the case of the $(-)$ branch of the solution, the event horizon formation for the shell focusing depends on the initial conditions.   Therefore, the SCS in this collapse scenario may not be accompanied by an event horizon.

The focus and scope of our paper herewith was to study the bound collapse $(k = 1)$ on DGP-GB brane. The study of a marginal bound model $(k = 0)$ was done in Ref.~\cite{Tavakoli:2020xgc}, which shows a similar result for the $(-)$ branch, in this sense that the formation of a naked or a black hole singularity depends on the initial condition. However, in the bound model $(k = 1)$, the $(+) $ branch of the solutions  does not violate the null energy condition in general, in contrast with the marginally bound system. Moreover, the $(+)$ branch in the bound collapse contains a characteristic solution that supports a  naked SCS, which is independent of the initial condition.

%%%%%%%%%%%%%%%%%%%%%%%%%%%%%%%%%%%%%%%%%%%%%%%%%%%%%%%%%%%%%%%%%%%%%%%%%%%%%%%%%%%%%%%

\section*{Acknowledgments}

\noindent P.V.M. and Y.T. acknowledge the the Portuguese agency
Funda\c{c}\~{a}o para a Ci\^{e}ncia e a Tecnologia (FCT)  Grants No. UIDBMAT/00212/2020 and No. UIDPMAT/00212/2020 at CMA-UBI. This article is based upon work from the Action CA18108 -- Quantum gravity phenomenology in the multi-messenger approach -- supported by the COST (European Cooperation in Science and Technology). 
The work of M. B. L. is supported by the
Basque Foundation of Science Ikerbasque. She acknowledges the support from the Basque government Grant No. IT956-16 (Spain) and from the Grant  No. PID2020-114035GB-100 funded
by Ministerio de Ciencia e Innovaci\'{o}n (MCIN) MCIN/AEI/10.13039/501100011033. She is also partially supported by the  Grant  No. FIS2017-85076-P, funded by MCIN/638 AEI/10.13039/501100011033 and by ``ERDF A way of making Europe.''

%The work of M.B.L. is supported by the Basque Foundation of Science Ikerbasque. She  acknowledges the support from the Basque government Grant No. IT956-16 (Spain) and from the Grant No. PID2020-114035GB-100 funded by MCIN/AEI/10.13039/501100011033. She is also partially supported by the Grant No. FIS2017-85076-P,  funded by MCIN/638 AEI/10.13039/501100011033 and by ``ERDF A way of making Europe''.

\appendix

\section{The cubic equation}

In this appendix, we review very briefly  a  cubic equation representing the Friedmann equations (\ref{Friedmannnb}) or  (\ref{Friedmannnb2}) and their possible solutions following the prescription provided in Ref.~\cite{Abramowitz:1965}.

Let us consider the general cubic equation:
\begin{eqnarray}
z^3 + a_2 z^2 + a_1 z + a_0 = 0 \ .
\label{cubic}
\end{eqnarray}
Let us define the variables as
\begin{eqnarray}
Q&=&\frac{1}{3}\left(a_1 - \frac{a_2^2}{3}\right) , \\
S&=& \frac{1}{6}\left(a_1a_2 -3a_0\right) - \frac{a_2^3}{27} \, ,
\end{eqnarray}
and let us also introduce the discriminant function $N$ as
\begin{eqnarray}
N= Q^3 + S^2 .
\end{eqnarray}
For $N>0$ there exists only one real root and a pair of complex conjugate roots; for $N=0$ there exists three real roots but at least two of them are equal; for $N<0$ there exists three real roots (irreducible case).

By defining the new variables
\begin{eqnarray}
S_1 &=& \left[S + \sqrt{N} \right]^{1/3} , \\
S_2 &=& \left[S- \sqrt{N} \right]^{1/3} ,
\end{eqnarray}
we can write the three solutions to the cubic equation (\ref{cubic}) as
\begin{subequations}
	\label{cubic-sol-gen}
	\begin{eqnarray}
		Z_1 &=& (S_1 + S_2) - \frac{a_2}{3}\, , \\
		Z_2 &=& -\frac{1}{2}(S_1 + S_2) - \frac{a_2}{3} + \frac{i\sqrt{3}}{2}(S_1-S_2) , \\
		Z_3 &=& -\frac{1}{2}(S_1 + S_2) - \frac{a_2}{3} - \frac{i\sqrt{3}}{2}(S_1-S_2) .
	\end{eqnarray}
\end{subequations}

\bibliography{Bibliography}

\end{document}